\documentclass[useAMS,usenatbib]{mnras}

\usepackage{graphics}
\usepackage{amssymb}
\usepackage{amsfonts}
\usepackage{wasysym}
\usepackage{epsfig}
\usepackage{breakurl}

\usepackage{longtable}
\usepackage{times}
\usepackage{mathptmx}
\usepackage[usenames,dvipsnames]{color}
\usepackage{url}

\title[The radio emission from KQ\,Vel]
{Discovery and origin of the radio emission from the multiple stellar system KQ\,Vel}
\author[P. Leto et al.]
{P. Leto,$^{1}$ \thanks{E-mail: paolo.leto@inaf.it}
L. M. Oskinova,$^{2}$ 
C. S. Buemi,$^{1}$ 
M. E. Shultz,$^{3}$
F. Cavallaro,$^{4,1}$ 
C. Trigilio,$^{1}$
\newauthor 
G. Umana,$^{1}$
L. Fossati,$^{5}$ 
I. Pillitteri,$^{6}$ 
J. Krti\v{c}ka,$^{7}$ 
R. Ignace,$^{8}$ 
C. Bordiu,$^{1}$ 
F. Bufano,$^{1}$
\newauthor 
G. Catanzaro,$^{1}$
L. Cerrigone,$^{9}$ 
M. Giarrusso,$^{10}$ 
A. Ingallinera,$^{1}$
S. Loru,$^{1}$
S. P. Owocki,$^{3}$
\newauthor 
K. A. Postnov,$^{11,12}$
S. Riggi,$^{1}$
J. Robrade,$^{13}$ 
F. Leone$^{14,1}$
\\
$^{1}$INAF -- Osservatorio Astrofisico di Catania, Via S. Sofia 78, I-95123 Catania, Italy\\ 
$^2$Institute for Physics and Astronomy, University Potsdam, D-14476 Potsdam, Germany\\ 
$^3$Department of Physics and Astronomy, University of Delaware, 217 Sharp Lab, Newark, Delaware, 19716, USA\\
$^{4}$The Inter-University Institute for Data Intensive Astronomy (IDIA), Department of Astronomy, University of Cape Town, Rondebosch, 7701, South Africa\\ 
$^5$Space Research Institute, Austrian Academy of Sciences, Schmiedlstrasse 6, A-8042 Graz, Austria\\ 
$^{6}$INAF -- Osservatorio Astronomico di Palermo, Piazza del Parlamento 1, I-90134 Palermo, Italy\\ 
$^7$Department of Theoretical Physics and Astrophysics, Masaryk University, Kotl\'{a}\v{r}sk\'{a} 2, CZ-611 37 Brno, Czech Republic\\ 
$^8$Department of Physics \& Astronomy, East Tennessee State University, Johnson City, TN 37614, USA\\ 
$^{9}$Joint ALMA Observatory, Alonso de C\'{o}rdova 3107, Vitacura, 8320000, Santiago, Chile\\  
$^{10}$Department of Physics and Astronomy, University of Florence, Largo Enrico Fermi 2, 50125, Firenze, Italy\\
$^{11}$Sternberg Astronomical Institute, M.V. Lomonosov Moscow University, Universitetskij pr. 13, 119234 Moscow, Russia\\
$^{12}$Kazan Federal University, Kremlevskaya Str 18, 42008, Kazan, Russia\\ 
$^{13}$Hamburger Sternwarte, University of Hamburg, Gojenbergsweg 112, D-21029 Hamburg, Germany\\ 
$^{14}$Dipartimento di Fisica e Astronomia, Sezione Astrofisica, Universit\'{a} di Catania, Via S. Sofia 78, I-95123 Catania, Italy
}
\begin{document}

\date{}

\pagerange{\pageref{firstpage}--\pageref{lastpage}} \pubyear{}

\maketitle

\label{firstpage}

\begin{abstract}
KQ\,Vel is a binary system composed of a slowly rotating magnetic Ap star with a companion of unknown nature. In this paper, we report the detection of its radio emission. We conducted a multi-frequency radio campaign using the ATCA interferometer (band-names: 16cm, 4cm, and 15mm). The target was detected in all bands. The most obvious explanation for the radio emission is that it originates in the magnetosphere of the Ap star, but this is shown unfeasible. The known stellar parameters of the Ap star enable us to exploit the scaling relationship for non-thermal gyro-synchrotron emission from early-type magnetic stars. This is a general relation demonstrating how radio emission from {stars with centrifugal magnetospheres}  is supported by rotation. Using KQ\,Vel's parameters the predicted radio luminosity is more than five orders of magnitudes lower than the measured one. The extremely long rotation period rules out the Ap star as the source of the observed radio emission. Other possible explanations for the radio emission from KQ\,Vel, involving its unknown companion, have been explored. A scenario that matches the observed features (i.e. radio luminosity and spectrum, correlation to X-rays) is a hierarchical stellar system, where the possible companion of the magnetic star is a close binary (possibly of RS\,CVn type) with at least one magnetically active late-type star. To be compatible with the total mass of the system, the last scenario places strong constraints on the orbital inclination of the KQ\,Vel stellar system.
\end{abstract}


%
%
%
\section{Introduction}
\label{sec_intro}

The magnetic field topology of early-type magnetic stars is commonly described by the simple 
yet highly successful
Oblique Rotator Model (ORM, \citealp{babcock49,stibbs50}).
In this model the magnetic field has a dipole dominated topology,
with the magnetic axis misaligned with respect to the stellar rotation axis. 
The stellar magnetic field (typical polar strength at the kG level, e.g. \citealp{shultz_etal19_490})
induces anisotropic chemical distributions at the stellar photosphere,
which are responsible for the observed variability \citep{krticka_etal07}.
The ORM explains the typical variability (photometric and spectroscopic) of such stars  
as a consequence of stellar rotation.
First discovered by \citet{ladstreet_borra78}, 
such stars may also be surrounded by a large-scale magnetosphere, 
filled with ionized material continuously lost from the stellar surface via a weak radiatively driven wind \citep{babel95, babel96}.

{The ability of the magnetosphere to sustain the confined plasma against gravitational
infall depends on a combination of field strength, plasma mass, and rotational speed
\citep{petit_etal13}.}
{Fast-rotating stars need strong magnetic fields to confine 
the centrifugally supported plasma to co-rotate with the star, 
leading to the formation of an extended Centrifugal Magnetosphere (CM).}
{The average rotation period  and  polar magnetic field strength of the stars with CMs
analyzed by \citet{petit_etal13} are $P_{\mathrm{rot}}\approx 3$ days and $B_{\mathrm p} \approx 6000$ G. The size of the CM}
is quantified by the Alfv\'{e}n radius ($R_{\mathrm A}$).
The Alfv\'{e}n radius is
where the magnetic field strength becomes no longer able to confine the plasma, setting
the extent of the magnetosphere.
In order to form a CM, the Alfv\'{e}n radius must 
be larger than the Kepler co-rotation radius ($R_{\mathrm K}$).
The photometric and spectroscopic variability 
from the CM has been modeled by the Rigidly Rotating Magnetosphere  (RRM) model
\citep{towsend_owocki05,towsend08,oksala_etal15,krticka_etal22,berry_etal22}.
Different observing diagnostic techniques have been taken into account
to discriminate between CMs and Dynamical Magnetospheres (DM), 
where in the latter the plasma falls back to the star owing to gravity.
In particular, the study of the H$\alpha$ emission features
has been explained by the presence of Centrifugal BreakOut (CBO) events \citep{ud-doula_etal08}
continuously occurring within the CMs \citep{owocki_etal20,shultz_etal20}. 
 
The magnetospheres of early-type magnetic stars are frequently sources of 
non-thermal radio emission 
produced by a population of relativistic electrons that, moving within the stellar magnetosphere,
radiate in the radio regime via the incoherent gyro-synchrotron emission mechanism 
\citep{drake_etal87,linsky_etal92,leone_etal94}.
Furthermore, a rapidly increasing number of early-type magnetic stars 
have also been discovered to be sources of strongly circularly polarized pulses (up to $\approx 100 \%$)
\citep{trigilio_etal00,
das_etal18,
das_etal19a,
das_etal19b,
leto_etal19,
leto_etal20,
leto_etal20b,
das_etal21}, 
that are produced by the electron cyclotron maser coherent emission mechanism \citep{trigilio_etal00}.
This coherent emission process has been explained as stellar auroral radio emission \citep{trigilio_etal11},
similar to the terrestrial auroral radio emission.
Such coherent emission, originating above the magnetic polar caps,
is intrinsically highly beamed, with the emission pattern mainly oriented perpendicular to the magnetic axis.
The maser emission is then usually visible within a limited range of stellar rotational phases close to the nulls of the effective magnetic curve.
In some cases the coherent emission is visible over a broad portion of stellar rotational phases.
This depends on a favorable stellar geometry \citep{leto_etal20b} or on the adopted (very low, typically a few hundred MHz) 
observing frequency \citep{das_chandra21}.

The gyro-synchrotron radio emission arising from the co-rotating magnetospheres of early-type magnetic stars
is modulated by stellar rotation \citep{leone91,leone_umana93}, 
with the amplitude of the radio emission variability 
being a function of the ORM geometry,
behavior that was successfully modeled by \citet{trigilio_etal04}.
In particular, stars showing large-amplitude magnetic curves are also characterized by the larger rotational variability
of their radio emission, i.e. see the cases of CU\,Vir (HD\,124224), HD\,37479 ($\sigma$\,Ori\,E), or HR\,7355 (HD\,182180)
that are characterized by large magnetic \citep{oksala_etal10,oksala_etal12, Rivinius_etal10, kochukhov_etal14}
and radio rotational variability \citep{leto_etal06,leto_etal12,leto_etal17}, 
and the case of $\rho$\,Oph\,C (HD\,147932) that is instead characterized 
by small-amplitude rotational variability of its radio emission \citep{leto_etal20b}.
Although exceptions exist,
i.e. HR\,5907 (HD\,142184) has a small-amplitude magnetic curve \citep{grunhut_etal12}
and a rotational variability of the radio emission (at $\nu \gtrapprox 20$ GHz) 
with amplitude which increase as the observing frequency increases
\citep{leto_etal18}.

Recently, it was empirically found that 
the radio luminosity for incoherent non-thermal radio emission from large-scale corotating magnetospheres
is nearly proportional to the square of the ratio between the unsigned magnetic flux 
($\Phi=B_{\mathrm p} R_{\ast}^2$, where $B_{\mathrm p}$ is the polar strength of the dipole-like magnetic field)
and the stellar rotation period ($P_{\mathrm{rot}}$), 
$L_{\nu,\mathrm{rad}} \propto (\Phi / P_{\mathrm{rot}})^2$ 
\citep{leto_etal21,shultz_etal22}.
{The underling physical mechanism supporting this empirical relation are continuously occurring CBO events.
In the case of a simple dipole-shaped magnetic field topology, 
the magnetic field strength ($B$) in the magnetic equatorial plane rapidly decreases outward
($B \propto r^{-3}$ as a function of the radial distance $r$). 
The centrifugal force acting on the co-rotating plasma breaks the magnetic field lines.
The magnetospheric region where CBO occurs is located at a well defined radial distance,
where the magnetic tension is no longer able to constrain the centrifugal force.
The resulting reconnection of the magnetic fields drives the acceleration of the local electrons 
that power the radio emission \citep{owocki_etal22}.
}
The existence of this general relation opens a new era 
in the radio study of magnetic stars surrounded by stable stellar magnetospheres.
In fact, the measurement of radio luminosity has proved to be a powerful tool 
for indirect estimation of some fundamental parameters of such stars.

In this paper we present new radio observations of KQ\,Vel, performed with the ATCA interferometer.
KQ\,Vel is a multiple stellar system. The brightest component is a well studied magnetic Ap star \citep{bailey_etal15},
whereas the nature of the companion is not yet clear.
The detection of the radio emission of KQ\,Vel
provides useful new information for advancing our
comprehension of the nature of
this enigmatic stellar system. The known properties of KQ\,Vel are summarized in Sec.~\ref{sec_kqvel}.
In Sec.~\ref{sec_radio_obs} we describe how the radio measurements were performed and
how these data have been analyzed.
The possible magnetospheric origin of the radio emission from KQ\,Vel is discussed in Sec.~\ref{sec_radio_magnetoshere}.
The scenario involving the radio emission arising from a possible degenerate companion
is discussed in Sec.~\ref{sec_radio_neutron_star}, where
the cases for both thermal and non-thermal origins 
have been analyzed.
In Sec.~\ref{sec_binary_star} the explanation of the radio emission observing features related to the possible existence of an active binary system
has been also taken into account.
The results of the analyses performed in this paper 
have been discussed in Sec.~\ref{sec_discussion}.
In Sec.~\ref{sec_conclusion} we proposed our conclusions
and possible further steps for the next investigations 
aimed to definitively unveil the nature of the KQ\,Vel stellar system.

\section{Summary on KQ\,Vel}
\label{sec_kqvel}

KQ\,Vel (HD\,94660) is a magnetic bright (visual magnitude 6.11) Ap star of the southern hemisphere
about $260$ Myr old \citep{kochukhov_bagnulo06}.
{Based on the early data of the third release of the Gaia mission \citep{gaia_dr3}
(data confirmed by the Gaia third release; \citealp{gaia_dr3_2022}),
\citet{bailer_etal21} estimates the photogeometric distances and locates}
KQ\,Vel at $160 \pm 8$ pc from Earth, which is compatible with the 
distance reported by Hipparcos ($d=150 \pm 20$ pc; \citealp{hipparcos}).
The first measurement of the magnetic field of KQ\,Vel was reported by \citet{borra_landstreet75}.
In the ORM paradigm, the stellar rotation induces
the variability of the projected magnetic field components integrated over the whole visible disk. 
In fact, the magnetic field measurements of KQ\,Vel 
have been used to estimate the stellar rotation period.
In particular, this Ap star is a long-period slow rotator. 
The first estimation of its rotation period, which was performed using magnetic field measurements, 
was $P_{\mathrm{rot}} \approx 2700$ d.
This period was obtained from the variation of the modulus of the mean magnetic field at the stellar surface \citep{landstreet_mathys00}.
This is in good accord with the period previously estimated by the photometric variability \citep{hensberge93}.
The period estimation has since been refined using magnetic field measurements covering a wider temporal baseline by \citet{landstreet_etal14}
($P_{\mathrm{rot}}=2800\pm 250$ d).
Finally, collecting new high sensitivity magnetic field measurements (typical errors a few tens of gauss), 
\citet{giarrusso_etal22} both confirmed the above period and improved on its precision
($P_{\mathrm{rot}}=2830\pm 140$ d).

The measured upper limit of the projected rotational velocity ($v \sin i < 2$ km s$^{-1}$)
and the modeling of the magnetic field variability
indicate that the stellar rotation axis of this magnetic Ap star is inclined by only a few degrees ($i_{\mathrm{rot}} \approx 16^{\circ}$)
with respect to the line of sight  
\citep*{bailey_etal15}.
Further, the measured effective magnetic field of KQ\,Vel is always with negative polarity, 
showing a small amplitude of rotational variability ($\approx 800$ G).
This means that KQ\,Vel is characterized by an ORM geometry 
where the negative stellar magnetic pole is always visible.
As a first-order approximation, 
the magnetic field topology of KQ\,Vel is described by a simple 
less tilted dipole (tilt angle with respect to the rotation axis $\beta \approx 30^{\circ}$)
with a polar magnetic field strength $B_{\mathrm p} \approx 7500$ G \citep*{bailey_etal15}.

This magnetic Ap star is also a member of a multiple stellar system.
The binary nature of KQ\,Vel was first reported by \cite{mathys_etal97}
revealing also that the orbital period is significantly shorter than the rotation period of the Ap star.
Multi-epoch high-resolution spectra evidenced a clear radial velocity variation (amplitude $\approx 35$ km s$^{-1}$) 
of $\approx 840$ d \citep*{bailey_etal15}.
Collecting a large number of additional spectra, the orbital period,
the mass function, and other orbital parameters have been constrained by \citet{mathys17}.
These orbital parameters are well in good agreement with values 
estimated by \citet{giarrusso_etal22}, the latter having slightly higher uncertainty.

High-quality visual spectra of KQ\,Vel did not evidence any clear spectral signatures able 
to characterize the nature of the companion of the bright magnetic Ap star \citep*{bailey_etal15}.
On the other hand, KQ\,Vel is also a bright X-ray source whose origin cannot be simply related 
to the visible Ap star \citep{oskinova_etal20}.
To explain how the X-ray emission from KQ\,Vel originates, different scenarios involving the unknown companion have been proposed.
\citet{oskinova_etal20} considered various scenarios, including a possibility that
the companion is a RS CVn-type binary consisting of late-type stars with high-level of coronal magnetic activity. 
This possibility was
discarded because typical activity signs (e.g. CaII spectral lines) have not been reported in the literature.
Finally, \citet{oskinova_etal20} favored a scenario involving the presence of
a hot shell surrounding a massive degenerate unseen companion 
(a propelling magnetic neutron star)
as responsible for the observed X-ray emission.
However,
the detection of a faint infrared source close to the bright Ap star, 
with short-period (2.1 d) photometric variability, and of flares
\citep{scholler_etal20} re-opened  the possibility that the companion of the Ap star is a binary system composed of a pair of
non-degenerate stars, with masses close to the mass of the Sun and
characterized by solar-like magnetic activity able to explain the measured X-ray emission level.

\section{Radio observations}
\label{sec_radio_obs}

The radio observations of KQ\,Vel were performed with the ATCA
(Australia Telescope Compact Array)\footnote{The Australia Telescope Compact Array is part of the Australia Telescope
National Facility which is funded by the Australian Government for
operation as a National Facility managed by CSIRO}
using the CABB wide-band backend, that allow observation of a frequency window $\approx$ 2 GHz wide.
The source was observed during the $\approx 12.5$ hours of visibility above the horizon limit.
We cyclically alternated three receivers: the first with central frequency band tuned to $\nu=2.1$ GHz (band name 16cm);
the second that is able to acquire two simultaneous bands tuned to $\nu=5.5$ and $\nu=9$ GHz 
(band name 4cm);  and the third (band name 16mm) with a setup 
that acquires two simultaneous contiguous bands tuned to $\nu=17$ and  $\nu=19$ GHz.
The available observing time (reduced for array setup and calibrations) allowed us to observe
KQ\,Vel at six different hour angles, for each band, 
which was enough to properly sample the uv-plane using the ATCA interferometer which has a linear array design.

\begin{table}
\caption[ ]{Observing log. Array configuration 6B; Code: C3379.}
\label{obs_log}

\begin{tabular}{@{~}l @{~~}c@{~~~~} c@{~~~~} c @{~~~}c @{~}}

\hline

Date&$<$UT$>$ & $\nu$ &Phase cal & Flux cal  \\
 &        & (GHz) &            & \\
\hline
2020-Oct-20 &22:30 &2.1 / 5.5 / 9 / 17 / 19 &$1104-445$  &  $1934-638$ \\
\hline

\end{tabular}

\end{table}

The phase calibrator PKS\,$1104-445$ was observed at all selected bands. 
The source $1104-445$ is a standard ATCA calibrator, that is ideal to calibrate the phases of the complex visibilities of KQ\,Vel
(distance from KQ\,Vel $3.38$ deg).
Within each individual scan, 
the observations were performed by cyclically switching the pointing of the target (KQ\,Vel) and of the phase calibrator.
Due to the time lost for phase calibrator observations, 
the effective time on source was about $150$ minutes per band.

The flux density scale was defined by observing the Seyfert Galaxy PKS\,$1934-638$,
which is the standard primary calibrator for the ATCA, 
used also to calibrate the frequency-dependent receiver response (bandpass calibration). 
The flux density uncertainty of the flux calibrator $1934-638$ is below $0.2 \%$ in all the observing bands.
Details regarding the observations are summarized in Table~\ref{obs_log}.

\begin{figure*}
\resizebox{\hsize}{!}{\includegraphics{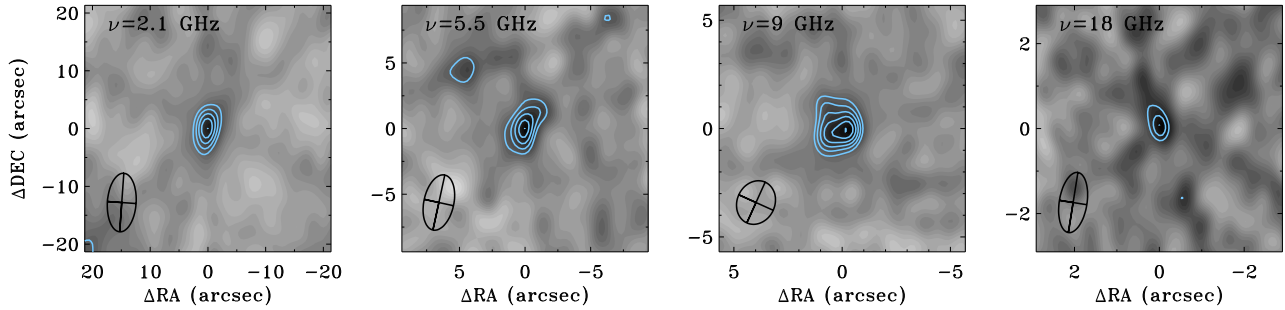}}
\caption{Maps performed at the sky position of KQ\,Vel at the indicated observing frequency.  
The light-blue contour levels are scaled in unit of the RMS of each map (values reported in Table~\ref{tab_flux}); 
the lower level is fixed at $3\sigma$.
The elliptical synthetic beams of the ATCA linear interferometer 
obtained using all the available observing scans are pictured in the left-bottom corner of each map.}
\label{radio_maps}
\end{figure*}

The data have been edited and calibrated using the software package
{\sc miriad}, which is the standard for ATCA measurements.
The bad data affected by strong RFI have been flagged (task {\sc blflag}).
The single observing scans performed at different hour angles allowed us to perform 
the cleaned maps (for all the observing bands) centered at the sky position of KQ\,Vel 
(tasks {\sc invert}, {\sc clean}, and {\sc restore}).
Cleaned maps  (2000 iterations) were obtained
at each observing frequency for both $I$ and $V$ Stokes parameters using the natural weighting.
The Stokes\,$I$ parameter is always positive and is the measure of the total intensity of the electromagnetic wave,
that is composed of
the two opposite circularly polarized components with right (RCP)
and left (LCP) rotation senses:
{Stokes\,$I=({\mathrm{RCP}}+\mathrm{LCP})/2$.
The Stokes\,$V =({\mathrm{RCP}}-\mathrm{LCP})/2$} 
instead measures the intensity of the circularly polarized radiation and
its sign is related to the dominant polarization state.
{The sign/orientation of the circularly polarized radiation follows the IAU/IEEE convention.
That is, the polarization plane of the incoming right-handed circularly polarized radiation is seen to rotate counterclockwise.
Conversely, the LCP circularly polarized radiation rotates clockwise.}

Due to the need to flag channels and scans 
corrupted by strong RFI, the effective observing bandpass and integration time were reduced, with the consequent increase of map noise.
The noise of the maps obtained in each band are listed in column 4 of Table~\ref{tab_flux}.
The measured map noise for the two bands available using the 4cm receiver 
was close (but worse) to the theoretical expected noises level ($\lessapprox 10$ $\mu$Jy/beam).
In any case, the total intensity (Stokes\,$I$) radio emission of KQ\,Vel was clearly detected in both bands.
The hard data flagging and the crowded field most significantly affected the noise level of the 16cm observations. 
The noise measured in the total intensity map is $\approx 30$ $\mu$Jy/beam, which is significantly 
higher than the expected value ($\approx 10$ $\mu$Jy/beam);
however, the target was safely detected (Stokes\,$I$ only) well above the $5 \sigma$ detection threshold.
The noise measured in the map of Stokes\,$V$ performed at 2.1 GHz 
is $\lessapprox 20$ $\mu$Jy/beam.
As expected, the noise measured in the Stokes\,$I$ map was significantly higher due 
to the large number of unpolarized bright sources present in the crowded field around KQ\,Vel.

\begin{table}
\caption[ ]{Measured radio fluxes and ATCA beams at each frequency.}
\label{tab_flux}

\begin{tabular}{@{~}c @{~~~}r  r c c@{~~~} r@{~}}

\hline
{$\nu$}       &{S}~~~~                &$\pi_{\mathrm c}$~~~~ &RMS             &FWHM  & PA\\
(GHz)         &{($\mu$Jy)}  &                           & ($\mu$Jy/beam)     &($^{\prime\prime} \times ^{\prime\prime}$) & (degree)\\
\hline
2.1              &{$201 \pm 47$ } & $\lessapprox14 \%^{\dag}$ & 28  &$10.16 \times 4.98~~$  & $-4.91$  \\
5.5              &{$86 \pm 18$ } & $ \lessapprox14 \%~~$ & 13               &$4.29 \times 2.28$  & $-12.14$ \\
9              &{$163 \pm 32$ }  & $ \lessapprox 9 \%~~$ &15               &$2.06 \times 1.74$  & $-24.30$\\
$18^{\ast}$ &{$94 \pm 23$ }  & $ \lessapprox 20 \%~~$ &18          &$1.43 \times 0.65$  & $-9.04$  \\

\hline

\end{tabular}

\begin{list}{}{}
\item[]{$^{\dag}$} Calculated using the noise measured on the Stokes\,$V$ map (20 $\mu$Jy).
\item[]{$^{\ast}$} Average of the two bands of the 15mm receiver.
\end{list}

\end{table}

At higher frequencies KQ\,Vel was not clearly detected in each single band available for the 15mm receiver.
Hence, to improve the signal to noise level
we combined the two bands centered at $\nu=17$ and  $\nu=19$ GHz.
After the averaging process, the central frequency is $\nu = 18$ GHz and
the corresponding map noise becomes comparable (or better) to
the noise levels theoretically expected 
for the maps performed at the two single bands of the 15mm receiver ($\approx 20$ $\mu$Jy/beam).
This enables us to report the detection ($5 \sigma$ detection level) of KQ\,Vel at this high frequency.

The target is seen as a point-like source at each frequency, except at $\nu=9$ GHz.
The radio maps performed at each observing frequency are shown in Fig.~\ref{radio_maps}.
The fluxes have been measured by fitting a bi-dimensional gaussian function at the sky position of KQ\,Vel,
with the fit performed using 
{the two-dimensional fitting tool implemented within the task {\sc imview} (``gaussfit'' button) of}
the {\sc casa} package.
Even if the source at 9 GHz seems slightly resolved (see Fig.~\ref{radio_maps}),
the deconvolution from the instrumental beam, computed by the bi-dimensional gaussian fitting procedure,
is not able the resolve the source, reporting the message that the 
``source may be (only marginally) resolved in only one direction''.
In any case, the faint flux level combined with the limited array spatial resolution
prevents reliable clues on the morphology of this radio source.

{The errors related to the measured flux densities of KQ\,Vel, in all bands, 
have been computed by adding in quadrature the noise of the maps (measured within regions far from background sources), 
the uncertainty related to the source fitting process, 
and 5\% of the fluxes to take into account calibration errors.}
The measured fluxes of KQ\,Vel, with the corresponding errors,
are listed in Table~\ref{tab_flux},
where the map noise at each observing frequency and
the corresponding upper limit of the circular polarization fraction ($\pi_{\mathrm c}=I/V$) have been also reported.
The measured peak intensity at 9 GHz
returned by the gaussian fit is $120 \pm 20$ $\mu$Jy/beam; that, within the related uncertainties, 
is compatible with the integrated flux listed in Table~\ref{tab_flux}. Note that,
in case of a point-like source the peak intensity value coincides with the flux of the unresolved source integrated over the instrumental beam.

The radio spectrum of KQ\,Vel has a slightly negative spectral slope characterized by a spectral index of $-0.2 \pm 0.2$. 
Finally, the data time series was split into two runs which were separately imaged
to search for possible time variability.
This process increases the noises of the maps, therefore
the faint radio signal of this source prevents a reliable analysis performed over shorter time scales.
Within these technical limitations, we did not detect significant time variability.

\section{Magnetospheric radio emission} 
\label{sec_radio_magnetoshere}

The average fluxes, reported in Table~\ref{tab_flux}, scaled 
at the distance of KQ\,Vel (see Table~\ref{star_par})
allows estimation of the radio spectral luminosity $L_{\nu,\mathrm{rad}}=4(\pm 2) \times 10^{15}$ erg s$^{-1}$ Hz$^{-1}$
at the mean frequency $<\nu> \approx 6.6$ GHz.
To explain the origin of radio emission from KQ\,Vel,
the most obvious scenario is to hypothesize that it arises from the 
magnetosphere of the bright early-type magnetic star.

The measured spectral index of the KQ\,Vel radio spectrum ($\alpha=-0.2 \pm 0.2$) is compatible with
the typical almost flat spectra of the non-thermal radio emission originating within the dipole dominated
magnetospheres of early-type magnetic stars,
as for example the strongly magnetized ($B_{\mathrm p} \approx 9.5$ kG; \citealp{shultz_etal19_490}) 
B2V star HR\,7355 (HD\,182180), whose rotationally averaged radio spectrum, 
covering a similar frequency range to that of the multi-frequency ATCA measurements of KQ\,Vel,
is characterized by a spectral index $\alpha \approx -0.1$
\citep{leto_etal17}.

In general,
the flat spectrum behavior of the gyro-synchrotron radio emission of the early-type magnetic stars 
covers quite a large spectral range, 
roughly tuned in the range $\approx1$--100 GHz \citep{leto_etal21}.
This is a consequence of the typical polar field strength of the sample of radio-loud stars ($\approx 1$--15 kG)
analyzed by \citet{leto_etal21}.
The observed spectral behavior of KQ\,Vel's radio emission is compatible with the typical behavior of the 
magnetospheric radio emission of early-type magnetic stars,
even if, on the basis of the ORM geometry and observed frequency range, 
KQ\,Vel is expected to have a small amplitude of radio emission rotational variability.
In any case, the extremely long rotation period of this magnetic star ($P_{\mathrm {rot}} \approx 2800$ d)
makes it impossible to detect any rotational modulation of its radio emission using current ATCA measurements.
The geometry of KQ\,Vel is also not favorable for the detection of possible coherent emission from this magnetic star.
In fact the stellar effective magnetic field does not invert sign (i.e., an absence of nulls),
furthermore the ATCA observations are not tuned at a very low observing frequency.
These unfavorable conditions were confirmed by the ATCA radio observations, 
which did not detect any evidence of circularly polarized radio emission from  KQ\,Vel (see Table~\ref{tab_flux}).

\begin{table}
\caption[ ]{Stellar parameters of KQ\,Vel.}
\label{star_par}

\begin{tabular}{l c c  }

\hline
$d$ (pc) & $160\pm 8$ & \citet{bailer_etal21}  \\
$P_{\mathrm{rot}}$ (d) & $2830 \pm 140$ &  \citet{giarrusso_etal22} \\

$T_{\mathrm{eff}}$ (K) & $11300 \pm 400$ &  \citet*{bailey_etal15} \\

$\log g$ (cgs) & $4.18 \pm 0.20$ &  \citet*{bailey_etal15} \\

$M_{\ast}$ (M$_{\odot}$) & $3.0\pm0.2$ &  \citet*{bailey_etal15} \\
$L_{\ast}$ (L$_{\odot}$) & $105 \pm 20$ &  \citet*{bailey_etal15} \\
$R_{\ast}$ (R$_{\odot}$) & $\approx 2.7 $ &  this paper \\
$B_{\mathrm{p}}$ (kG) & $\approx7.5$ &  \citet*{bailey_etal15} \\
$i_{\mathrm{rot}}$ (degree) & $\approx 16$ &  \citet*{bailey_etal15} \\
$\beta$ (degree) & $\approx 30$ &  \citet*{bailey_etal15} \\

\hline
\multicolumn{3}{c}{Orbital parameters \citep{mathys17}}\\
\hline
 {$P_{\mathrm{orb}}$ (d)} & $848.96 \pm 0.13$ &   \\
 $T_{0}$ (HJD) & $2\,445\,628.6 \pm 1.7$ &   \\
 $e$ & $0.4476 \pm 0.0049$ &   \\
 $V_0$ (km s$^{-1}$)  & $18.53 \pm 0.06$ &   \\
 $\omega$ (degree) & $264.5 \pm 0.8$ &  \\
 $K$ (km s$^{-1}$) & $17.94 \pm 0.11$ &   \\
 $f(M)$ (M$_{\odot}$) & $0.3631 \pm 0.0075$ &  \\
 $a \sin i$ (AU) & $1.25 \pm 0.01$ &   \\

\hline

\end{tabular}

\end{table}

The behavior of the radio emission from the early-type magnetic stars 
has been definitively quantified by the empirical scaling relationship 
between the radio spectral luminosity and the magnetic flux rate:
$L_{\nu,\mathrm{rad}} {\propto} (\Phi/ P_{\mathrm{rot}})^2 = {B_{\mathrm p}^2 R_{\ast}^{4}} / {P_{\mathrm {rot}}^2}$,
found by \citet{leto_etal21} and confirmed by \citet{shultz_etal22}. 
This empirical relationship is the consequence of the physical mechanism supporting the 
non-thermal acceleration able to produce the electrons responsible 
for the magnetospheric radio emission of early-type magnetic stars.
It was recently demonstrated by \citet{owocki_etal22}
that the power provided by centrifugal breakout events, continuously occurring within the stellar magnetosphere,
is directly related to the power of the radio emission.
From this point of view, the CBOs are non-random events.
Breakouts occur continuously in a well-constrained magnetospheric region,
where the resulting reconnection of the magnetic fields drives the acceleration of the local electrons.
The CBO luminosity
is related to the stellar parameters by the relation \citep{owocki_etal22}:
\begin{displaymath}
L_{\mathrm{CBO}} = \frac {B_{\mathrm p}^2 R_{\ast}^3} {P_{\mathrm{rot}}} \times W {\mathrm {~~(erg \, s^{-1}),}}
\end{displaymath}
with $B_{\mathrm p}$ in gauss, $R_{\ast}$ in centimeters, and $P_{\mathrm{rot}}$ in seconds.
In the above relation $W=2 \pi R_{\ast} / P_{\mathrm{rot}} \sqrt{G M_{\ast}/R_{\ast}}$
is the dimensionless critical rotation parameter calculated as the ratio between the equatorial stellar velocity 
and the corresponding orbital velocity,
defined by the gravitational law at the stellar equator
($G = 6.67408 \times 10^{-8}$ cm$^3$ g$^{-1}$ s$^{-2}$ is the gravitational constant).
{Once the explicit relation of $W$ is substituted within the $L_{\mathrm{CBO}}$ definition,
we find that $L_{\mathrm{CBO}} \propto {B_{\mathrm p}^2 R_{\ast}^{4.5} M_{\ast}^{-0.5}} / {P_{\mathrm {rot}}^2} \propto L_{\nu,\mathrm{rad}}$,
which in practice is almost the same relation empirically found, except for the term related to the mass,
that in any case has negligible effect.
}
The maximum possible value {of $W$} is $W=1$, 
which, due to magnetic braking,  would have to be at the beginning of the star's life; 
however no magnetic star has ever been found with $W \gtrapprox 0.5$ \citep{shultz_etal19_490}. 
In fact, this parameter progressively decreases  
as the star loses angular momentum throughout its life \citep{keszthelyi_etal19,keszthelyi_etal20}.

Once the radio spectral luminosity ($L_{\nu,\mathrm{rad}}$) is integrated over a 100 GHz 
wide frequency range, in which the radio spectrum of a typical early-type magnetic star 
can be reasonably assumed to be flat  \citep{leto_etal21},
the efficiency of the CBO process supporting radio emission is $\approx 10^{-8}$ \citep{owocki_etal22}.
Therefore, the radio and the CBO luminosities are simply related  by the relation \citep{owocki_etal22}: 
$L_{\mathrm{rad}} \approx 10^{-8} L_{\mathrm{CBO}}$.
{In the case of a flat spectrum, the radio spectral luminosity is related to 
the power radiated over a frequency range $10^{11}$ Hz wide by the simple relation:
$L_{\mathrm{rad}} {\mathrm{(erg\, s^{-1})}}=10^{11}{\mathrm{(Hz)}} \times L_{\nu,\mathrm{rad}}{\mathrm{(erg\, s^{-1}\, Hz^{-1})}}$.
As evidenced by \citet{leto_etal21}, 
the gyro-synchrotron radio emission of the early type magnetic stars fade outside such a flat spectrum region.
}

The relationship reported above 
can also be generalized in cases of radio sources characterized by 
radio spectra covering frequency ranges narrower than 100 GHz,
such as Jupiter's non-thermal radio spectrum that is almost flat only below $\nu \lessapprox 1$ GHz \citep{de_pater_dunn03},
depending on the lower strength of the Jovian magnetic field ($\approx 4.17$ G; \citealp{connerney_etal18})
compared to the case of the early-type magnetic stars.
{To compare the power provided by centrifugal breakout with the radio spectral luminosity,
which is the unique observable available in cases of single-frequency radio measurements,
the relationship is:
\begin{equation}
\frac   {L_{\mathrm{CBO}}} {L_{\nu, \mathrm{rad}}}  \approx 10^{19}\, \mathrm{Hz,}
\label{eq_ll2}
\end{equation}
\noindent
{which is derived from the relation $10^{-8} L_{\mathrm{CBO}}=10^{11}L_{\nu,\mathrm{rad}}$.}
This is formally similar to the G\"{u}del-Benz relationship between X-ray luminosity and radio spectral luminosity
characterizing a large range of stellar types with active coronae \citep{guedel_benz93,benz_guedel94}.}

\begin{figure}
\resizebox{\hsize}{!}{\includegraphics{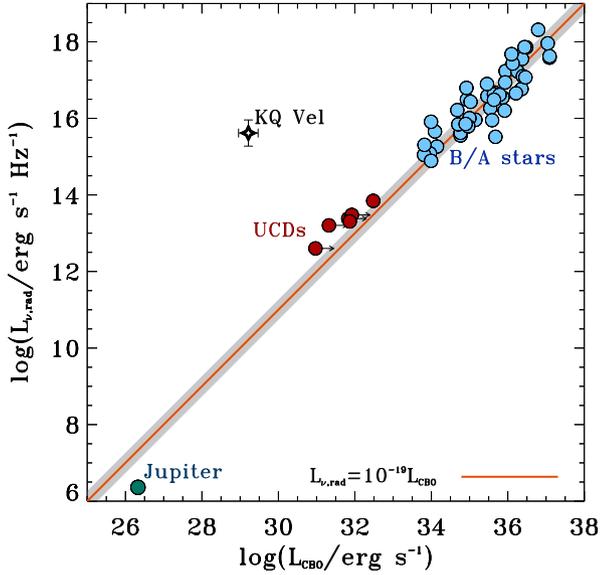}}
\caption{{
$L_{\nu,\mathrm{rad}} / L_{\mathrm{CBO}}$ diagram.
The orange solid line represents the 
generalized scaling relationship \citep{owocki_etal22} between the radio spectral luminosity (erg s$^{-1}$ Hz$^{-1}$) 
and the power provided by centrifugal breakouts (erg s$^{-1}$); see text.
This relation describes the common behavior of the non-thermal radio emission 
of stars and planets surrounded by large scale dipole-like co-rotating magnetospheres. 
The light-blue bullets represent the sample of early-type magnetic stars well studied at the radio regime \citep{leto_etal21,shultz_etal22}.
The grey region {around the orange line shows the statistical error of the B/A sample.}
The red bullets refer to the UCDs 
(the arrows refer to the cases with only estimates of the lower limit of the polar magnetic field strength).
The green bullet locates the planet Jupiter.
In the diagram
the location of the early-type magnetic star member of the KQ\,Vel system
is shown as a diamond symbol.
}}
\label{fig_lum_flum}
\end{figure}

The magnetic, radio-loud early-type stars 
show the common behavior described by the above relationship, pictured in Fig.~\ref{fig_lum_flum},
where the two samples of radio detected stars collected by \citet{leto_etal21} 
and by \citet{shultz_etal22} are also shown.
In Fig.~\ref{fig_lum_flum}, the cases of the non-thermal radio
emission of the planet Jupiter and of the fully convective Ultra Cool Dwarf stars,
already analyzed by \citet{leto_etal21}, are also shown. 
Note, that the non-thermal incoherent radio emission of the UCDs has been modeled taking into account
the existence of a plasma source responsible for the surrounding stellar environment
having a central symmetry \citep{ravi_etal11},
like the case of the early-type magnetic stars.

Note that an explanation of the slight discrepancy of Jupiter with respect to the universal law 
reported in Fig.~\ref{fig_lum_flum} was provided by \citet{owocki_etal22}.
This is related to the non-central location within the Jovian magnetosphere
of the plasma source: the volcanic moon Io that
deposits plasma in an inhomogeneous and anisotropic way into the
Io plasma torus.

{The main stellar parameters of the magnetic star of the KQ\,Vel system  
are reported in Table~\ref{star_par}.
The stellar radius is obtained scaling the value given by \citet*{bailey_etal15}
at the distance of KQ\,Vel reported in the third release of the Gaia mission \citep{bailer_etal21}.
In practice, the stellar radius provided by \citet*{bailey_etal15} has been corrected by a factor 1.07, 
which is the ratio between the distance of KQ\,Vel provided by Gaia and the distance from Hipparcos ($160/150 \approx 1.07$).
This correction leaves the angular extension of the star unchanged, making all the other stellar parameters reliable.
We note that the ORM parameters determined by \citet*{bailey_etal15} 
were derived purely from modeling the star's magnetic field measurements, 
and as such are insensitive to any change in the assumed stellar radius.
}
Using the stellar parameters listed in Table~\ref{star_par},
the CBO power of the magnetic star 
is then $L_{\mathrm{CBO}}\approx 1.6 \times 10^{29}$ erg s$^{-1}$, equivalent to
$\log{L_{\mathrm{CBO}}} \approx 29.2$.
The corresponding expected value of the radio luminosity of KQ\,Vel,
estimated using the scaling relationship for radio emission summarized by Eq.~\ref{eq_ll2},
predicts a luminosity level of about $1.6 \times 10^{10}$ erg s$^{-1}$ Hz$^{-1}$ 
($\log {L_{\nu, \mathrm{rad}}} \approx 10.2$),
which is substantially below the measured value of $\approx 4 \times 10^{15}$ erg s$^{-1}$ Hz$^{-1}$, or 
$\log {L_{\nu, \mathrm{rad}}} \approx 15.6$.
The location of KQ\,Vel  
in the ${L_{\nu, \mathrm{rad}}} / {L_{\mathrm{CBO}}}$ diagram
is not in accordance with the prediction of Eq.~\ref{eq_ll2}
(see Fig.~\ref{fig_lum_flum}).
Further, the above predicted magnetospheric spectral luminosity
is even lower than KQ\,Vel's photospheric contribution at $\nu=6.6$ GHz, which is about
$2.1 \times 10^{11}$ erg s$^{-1}$ Hz$^{-1}$, a value
estimated assuming the stellar photosphere radiates like a black body:
$L_{\nu}(T_{\mathrm{eff}})=4 \pi R_{\ast}^2 \times B_{\nu}(T_{\mathrm{eff}})$ erg s$^{-1}$ Hz$^{-1}$ 
(with the stellar radius given in centimeters).
The spectral flux density of the black body in the Rayleigh-Jeans approximation is
$B_{\nu}(T_{\mathrm{eff}})=2 \pi k_{\mathrm B} T_{\mathrm{eff}} \frac{\nu^2}{c^2}$ erg s$^{-1}$ cm$^{-2}$  Hz$^{-1}$
(with $k_{\mathrm B}=1.38 \times 10^{-16}$ erg K$^{-1}$ Boltzmann constant and $c=2.998 \times 10^{10}$ cm s$^{-1}$ speed of the light).

The enormous discrepancy between the measured radio spectral luminosity and 
the value predicted by the universal law (Eq.~\ref{eq_ll2}),
differing by $\approx 5$ dex, 
makes it hard to assign the radio emission from KQ\,Vel to being
magnetospheric in origin from the magnetic Ap star member of the KQ\,Vel system. 
The very long stellar rotation period places
the Kepler co-rotation radius for this Ap star at $R_{\mathrm K} \approx 450$ stellar radii
(obtained equating the centrifugal and the gravitational accelerations: 
$R_{\mathrm{K}} (2 \pi / P_{\mathrm{rot}})^2 = G M_{\ast} / R_{\mathrm{K}}^2$, 
estimated using the parameters listed in Table~\ref{star_par}).
Such a high value of $R_{\mathrm K}$ is likely larger than the Alfv\'{e}n radius.
The values of $R_{\mathrm A}$ calculated for a large sample of early-type magnetic stars 
are typically lower than 100 stellar radii
\citep{shultz_etal19_490}.
The estimated upper limit for $R_{\mathrm A}$ has been derived for stars hotter than the Ap star component of KQ\,Vel, 
typically $T_{\mathrm {eff}} \gtrapprox 15$ kK \citep{shultz_etal19_485}.
Using stellar parameters listed in Table~\ref{star_par},
we obtain $R_{\mathrm A} \approx 106$ R$_{\ast}$
adopting a wind mass-loss rate $\approx 10^{-11.9}$ M$_{\odot}$ yr$^{-1}$
using the recipe of \citet{vink_etal01},
or $R_{\mathrm A} \approx 415$ R$_{\ast}$ using
the mass-loss rate of $\approx 10^{-14.3}$ M$_{\odot}$ yr$^{-1}$ provided
by \citet{krticka14}. 
Either of these values for $R_{\mathrm A}$ are lower than $R_{\mathrm K}$.
Further, the adopted method overestimates the value of $R_{\mathrm A}$.
{In fact, the above analysis only takes into account the radiative wind,
where the Alv\'en radius is simply related to the wind confinement parameter $\eta=B_{\mathrm{eq}}^2 R_{\ast}^2 / \dot{M} v_{\infty}$ 
\citep{ud-doula_etal02} 
by the simple relation $R_{\mathrm A} \propto \eta^{1/4}$ \citep*{ud-doula_etal08,ud_doula_etal14}.}
The terminal wind velocity was assumed to be $v_{\infty}=500$ km\,s$^{-1}$, as adopted by \citet{oskinova_etal20}. 
But, as recently demonstrated \citep{leto_etal21}, the centrifugal effects locate the Alfv\'en radius still closer to the star.

The condition $R_{\mathrm A} < R_{\mathrm K}$
is unfavorable for the generation of an extended CM \citep{petit_etal13},
{this also implies unfavorable conditions for triggering
the CBO events able to supply the electrons necessary for
the magnetospheric radio emission.
The low radio emission level of KQ\,Vel predicted by the Eq.~\ref{eq_ll2}
indicates absence or negligible CBO events occurring 
within the magnetosphere of this slowly rotating magnetic star.
This is in accordance with 
the absence of a CM surrounding the star.
This is likely surrounded only by a low-density DM magnetosphere.
The measured radio luminosity of KQ\,Vel is therefore totally inconsistent with
the scaling relationship prediction. In fact no radio emission is
expected due to the absence of a CM,}
further casting doubt on the Ap star as the origin of the radio emission.

\section{Radio emission from a propelling NS} 
\label{sec_radio_neutron_star}

A magnetospheric origin for the measured KQ\,Vel radio emission was rejected in Sec~\ref{sec_radio_magnetoshere}.
In the following sections we analyze other scenarios that may reveal where the radio emission comes from.

KQ\,Vel is characterized by a high X-ray emission level.
The measured X-ray luminosity is $L_{\mathrm X} \approx 2 \times 10^{30}$ erg s$^{-1}$.
Model fitting of the X-ray spectrum of KQ\,Vel 
is compatible with thermal X-ray emission produced by hot electrons with a temperature higher than
$T\approx 20$ MK and an emission measure $EM=n_{\mathrm e}^2 V  \approx 5 \times 10^{52}$ cm$^{-3}$,
or with a thermal component combined with non-thermal photons 
following a power-law distribution \citep{oskinova_etal20}.

To explain the origin of the X-ray emission from KQ\,Vel,
a scenario involving a degenerate objects was proposed by \citet{oskinova_etal20}. 
If the unknown companion of the early type magnetic star is a magnetized NS of about 1.5 M$_{\odot}$,
the ionized material continuously lost by the radiative wind of the Ap star is gravitationally captured by the degenerate companion.
In this case, the NS magnetic field forces the plasma to corotate giving rise to a ``propeller effect''
\citep{illarionov75}, which can explain the origin of the hot emitting X-ray plasma.

\begin{figure}
\resizebox{\hsize}{!}{\includegraphics{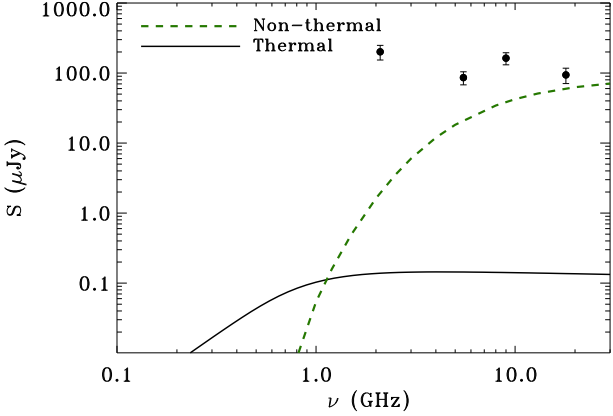}}
\caption{Thermal radio spectrum produced by the free-free emission mechanism
arising from the hot ionized shell surrounding a propelling NS (solid line).
The adopted model parameters are summarized in Sec.~\ref{sec_term_spectrum}.
The gyro-synchrotron radio spectrum of the NS magnetosphere (green dashed line) 
has been calculated assuming a spectral index $\delta=2$ for the power law of the non-thermal electron energy distribution
and with a number density $n_{\mathrm {rel}}=n_{\mathrm e}/10$.
The adopted source geometry is discussed in Sec.~\ref{sec_nonterm_spectrum}.}
\label{spe_free_free}
\end{figure}

\subsection{Thermal radio emission} 
\label{sec_term_spectrum}

\begin{figure}
\resizebox{\hsize}{!}{\includegraphics{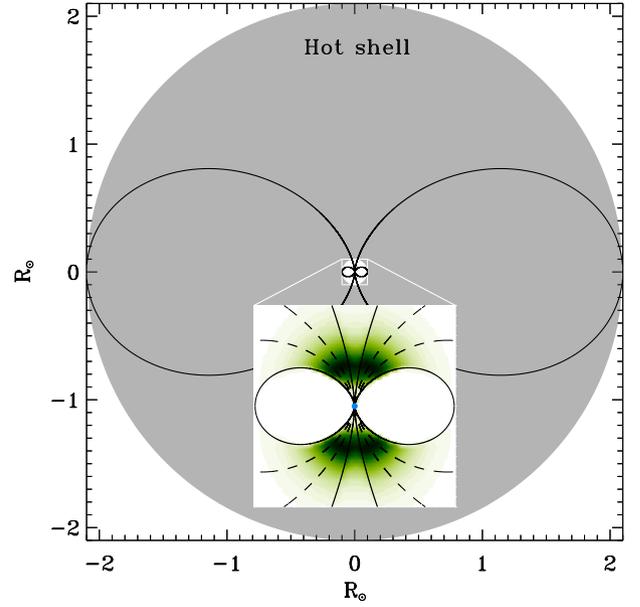}}
\caption{Synthetic brightness spatial distribution of the gyro-synchrotron radio emission (calculated at $\nu=30$ GHz)
from the magnetosphere of the magnetic neutron star hypothesized to explain the X-ray emission from KQ\,Vel. 
The X-rays originate from a large thermal hot shell surrounding the propelling NS (grey region).
The blue dot in the inset represents the magnetic NS (size not to scale).
The zoomed area ($0.2 \times 0.2$ R$_{\odot}$) 
shows that the brighter radio-emitting regions are mainly located at $\approx 0.04$ R$_{\odot}$ from the NS.
}
\label{scenario_ns}
\end{figure}

To reproduce the X-ray emission from KQ\,Vel,
the hot ionized material is trapped within a shell, 
with inner radius $R_{\mathrm {in}}=0.1$ R$_{\odot}$ and outer radius $R_{\mathrm {ext}}=2.2$ R$_{\odot}$,
centered on the propelling NS (see \citealp{oskinova_etal20} for details).
Using the emission measure of KQ\,Vel, 
following the relation 
$n_{\mathrm e} =\sqrt{EM / V_{\mathrm {shell}}}$
(where $V_{\mathrm {shell}}$ is the volume of the shell),
we derive an average electron density $n_{\mathrm e} \approx 1.8\times 10^9$ cm$^{-3}$ for this shell structure.

To check if the hot thermal plasma of the shell producing X-rays
might also explain the measured radio emission from KQ\,Vel,
we calculated the radio spectrum of this plasma shell.
The model for the free-free emission from thermal electrons trapped within a spherical shell-like region 
was first proposed by \citet{umana_etal08}, 
successfully applied to reproduce the radio spectra of planetary nebulae \citep{cerrigone_etal08},
and has also been used to study the radio emission from ultra compact HII regions \citep{leto_etal09}.

Once the  
density and temperature of the thermal electrons (assumed to be homogenously spatially distributed) and the shell size have been assigned,
the model is able to compute the radio emission produced by the free-free emission mechanism 
within a desired frequency range.
The adopted model parameters coincide with the physical and geometrical parameters derived to explain the X-ray emission from KQ\,Vel.
Those, for better clarity, are here summarized:
the geometry of the source is a shell having internal and external radii of respectively 0.1 and 2.2 R$_{\odot}$; 
the radiating electrons (with a temperature equal to 20 MK) are assumed to be homogeneously spatially distributed,
with an average density of $1.8 \times 10^9$ cm$^{-3}$.

The radio spectrum of KQ\,Vel due to the thermal emission of the hot X-ray plasma is pictured in Fig.~\ref{spe_free_free}.
The calculated emission level of the radio emission is too low ($\approx 3$ dex) to be able to reproduce the observed level.
This demonstrates that the simple thermal emission of the 
ionized shell surrounding a NS cannot explain the origin of the radio emission from KQ\,Vel.

\begin{table*}
\caption[ ]{Coronal parameters for the calculation of the radio spectrum using the core-halo model.}
\label{coha_par}
\begin{tabular}[]{c c c c ccc c ccc}
\hline
                           &~~~~~~        &       &               &\multicolumn{3}{c}{Core}              &~~~~~~  &\multicolumn{3}{c}{Halo}\\
                                                \cline{5-7}                                                                              
                                                                                                                                                        \cline{9-11}      
 
 Radius of the active star  &  &Size      &         &Radius   & $B$         & $n_{\mathrm{rel}}$       &     &Thickness  & $B$        & $n_{\mathrm{rel}}^0$  \\
                                                                                                                                                         
 (R$_{\odot}$)                  &  &               &          &(R$_{\ast}$)  &(G)     &(cm$^{-3}$)                   &  &(R$_{\ast}$)&(G)       &(cm$^{-3}$)                \\
 \hline

                                     &      &small           &      &0.4     &75                 &$2\times 10^{7}$            &  & 1      &7                 &$3\times 10^{7}$                \\     
1                                    &     &medium       &  &0.68     &220                 &$4\times 10^{5}$            &  & 3      &10                 &$2\times 10^{6}$                \\     
                                      &     &large            &  &0.8     &350                 &$1\times 10^{5}$            &  &7      &10                 &$7\times 10^{5}$                \\     
& & & & & & & & & \\
        
                               &       &small           &  &0.25     &95                 &$6\times 10^{6}$            &  &0.5     &7                 &$6\times 10^{6}$                \\    
2                              &      &medium         &  &0.35    &240                 &$3\times 10^{5}$            &  & 1     &10                 &$1\times 10^{6}$                \\    
                                &     &large &  &0.45     &400                &$5\times 10^{4}$            &  &3     &10                 &$2\times 10^{5}$                \\    
& & & & & & & & & \\
         
                             &       &small             &  &0.08     &55                 &$1\times 10^{8}$            &  &0.5      &7                 &$7\times 10^{5}$                \\             
4                           &        &medium          &  &0.15     &210                 &$7\times 10^{5}$            &  &1     &10                 &$1\times 10^{5}$                \\             
                             &      &large   &  &0.18     &310               &$2\times 10^{5}$            &  &2      &10                &$5\times 10^{4}$                \\             
\hline
                                                                                                                                                                      
\end{tabular}

\end{table*}

\subsection{Non-thermal radio emission} 
\label{sec_nonterm_spectrum}

The X-ray spectrum  of KQ\,Vel is also compatible with the presence of a non-thermal component. 
This makes plausible the existence of 
non-thermal electrons within the hot plasma shell surrounding the magnetized NS.
Following \citet{oskinova_etal20},  
a dipole magnetic moment of $\mu =3 \times 10^{30}$ G cm$^{3}$ was derived
for the propelling NS, typical for neutron stars,
corresponding to a polar magnetic field strength of $6 \times 10^{12}$ G,
obtained using the simple dipole relation at the distance of 10 km, the typical radius of a neutron star.
The corresponding average magnetic field strength, within the hot shell
responsible for the X-ray emission, is $<B> \approx 0.4$ G.
As discussed in the appendix of \citet{oskinova_etal20},
magnetic reconnection events can occur within the hot shell surrounding the NS,
with consequent generation of a non-thermal electron tail having a density $\approx 10\%$
of the ambient hot thermal plasma.
These relativistic electrons can power a non-thermal emission mechanism capable 
of producing detectable radio emission at the analyzed spectral range.

The harmonic numbers where the gyro-synchrotron emission mechanism efficiently works
are $10 \lessapprox \nu/\nu_{\mathrm{B}} \lessapprox 100$ \citep{dulk85,guedel02}, 
where $\nu_{\mathrm{B}}=2.8 \times 10^{-3} (B/{\mathrm G})$ GHz is the local gyro-frequency.
Within spatial regions characterized on the average by a magnetic field strength of about $0.4$ G,
the harmonic number is higher than 100 already at $\nu \approx 0.1$ GHz.
Hence,
to produce an order of magnitude estimate of the non-thermal incoherent radio emission 
from the magnetosphere of the NS, we assume that all non-thermal electrons produced within the hot shell
fall within spatial regions close to the NS, where the local magnetic field strength is high
enough to produce significant radio emission falling within the spectral range analyzed in this paper.

The calculation of the radio spectrum arising from the NS magnetosphere 
has been performed using the procedures developed to calculate the radio spectrum
from dipole-like magnetospheres surrounding early-type magnetic stars
\citep{trigilio_etal04,leto_etal06}.
The assumed energy spectrum of the relativistic electrons is a power law ($\propto E^{\delta}$),
where the spectral index is fixed ($\delta=2$) and the low-energy cutoff is assumed to be 10 keV,
whereas the high energy cutoff is fixed at 500 MeV.
In any case, in the frequency range $\approx 1$--100 GHz,
electrons with higher energy do not contribute to the non-thermal radio emission. 
The number density of the relativistic electrons is $n_{\mathrm{rel}}=0.1 \times n_{\mathrm e}=1.8 \times 10^8$ cm$^{-3}$.
These non-thermal electrons homogeneously fill the magnetospheric regions crossed by the magnetic field lines that 
cross the magnetic equator at distances larger than $R_{\mathrm{in}}=0.1$ R$_{\odot}$.

The calculated spectrum is pictured in Fig.~\ref{spe_free_free} using the dashed green line.
In Fig.~\ref{scenario_ns} the synthetic brightness spatial distribution of the radio emission
from the magnetosphere of the NS is shown.
It is clear that the radio emission would have to originate from regions located deep inside the propelling 
hot shell filled by thermal plasma.
The absorption effects suffered by the 
radio radiation traversing the hot shell 
have been taken into account.
The high-temperature thermal plasma of the hot shell
is optically thin at the higher radio frequencies,
whereas its absorption effects are not negligible at the lower frequencies.
For example, at the lowest frequency analyzed here ($\nu=2.1$ GHz)
the optical depth is $\tau_{\nu}\approx 0.23$, becoming larger than 1 at $\nu \lessapprox 1$ GHz.
These frequency-dependent absorption effects significantly affect the shape of the calculated non-thermal radio spectrum.
The comparison between measured and calculated radio emission shows that at
higher frequencies the theoretical emission level is close to the observed one,
whereas at the lower frequency side of the radio spectrum a clear discrepancy is evident.

The calculations performed here indicate that non-thermal radio emission is a
favorable mechanism able to produce significant (detectable) radio emission 
at the higher radio frequencies analyzed.
On the other hand, the hypothesized thermal plasma shell supporting the X-ray emission from KQ\,Vel
has a not-negligible optical depth for the radio waves produced at lower frequencies,
casting doubt on non-thermal emission from a propelling neutron star as the mechanism responsible for the measured radio emission. 
This shows that neither thermal nor non-thermal emission from a degenerate companion can reproduce KQ\,Vel's radio spectrum.

\section{Radio emission from a late type companion}
\label{sec_binary_star}

Among the various scenarios explored by \citet{oskinova_etal20}
to explain the high X-ray emission level of KQ\,Vel,
the possible existence of an active close binary orbiting around the Ap star was also considered.
But, this hypothesis was discarded due to the lack of evidence of the typical  spectral signatures 
of chromospheric magnetic activity \citep*{bailey_etal15}.
On the other hand, the evidence of photometric variability and flares \citep{scholler_etal20}
motivates reconsideration of the existence of a late type companion.

\begin{figure*}
\resizebox{\hsize}{!}{\includegraphics{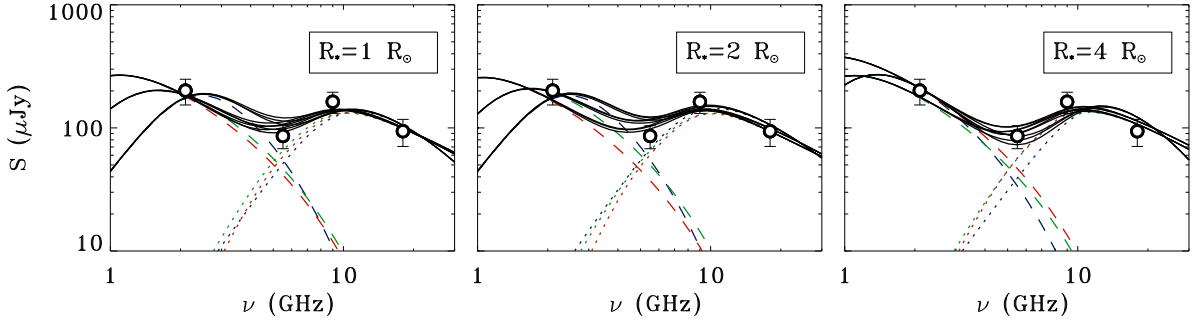}}
\caption{KQ\,Vel observed spectrum (open circles) with superimposed the radio spectra
calculated using a core-halo model for the gyro-synchrotron emission mechanism
from a stellar corona characterized by solar like magnetic activity.
Different model parameters have been explored.
The ``halo" spectrum (dashed lines) arises from extended regions permeated by
low strength magnetic field and dominating at the lower frequencies.
The radio spectrum from the ``core" (dotted lines) originates from a
more compact region with higher magnetic field strength, that dominates at the higher frequency range.
To distinguish the radio spectra calculated using coronal components of different sizes,
the corresponding spectra have been displayed using different colors
(the small, medium and large sizes are shown with blue, green, and red lines respectively).
The total spectra obtained from combining each component have been pictured using the solid lines.
Three different radii for the active star have been assumed for model calculations.}
\label{spe_late_star_nonterm_gyro_syn}
\end{figure*}

Chromospheric and coronal Solar-like magnetic activity is common in stars with deep
convective zones.
In the radio regime, coronal magnetic activity was recognized in the form of 
non-thermal radio emission characterized by stable emission levels 
and long-lasting active phases, with enhanced emission and flaring activity
\citep{trigilio_etal98}.
Such non-thermal radio emission was mostly detected in fast-rotating stars,
as confirmed by the correlation between the coronal radio emission level and the stellar rotation period
\citep{mutel_lestrade85}. This could be due to the dynamo mechanism amplified by rapid rotation.
In particular, in the case of late-type stars that are components of close binaries,  
the tidal orbital interaction makes these stars rotate faster than single stars of similar spectral types.
As an example, the basal emission level of the non-thermal coronal emission
from the large binary $\alpha$\,Cen\,A and B, 
the closest visible stars to Earth,
is below the thermal chromospheric radio emission of both solar type stars \citep{trigilio_etal18}.
As expected, the stellar rotation periods are not affected by their large orbital separation,
hence, the dynamo mechanism of the two stars 
does not seem to be amplified.
KQ\,Vel evidenced photometric variability with a period of 2.1 days,
much shorter than the 2800 d rotation period of the 
bright Ap star \citep{scholler_etal20}. 
The observed periodicity is likely of rotational origin.
This period is significantly shorter ($\approx 1$ dex) than the Sun's period,
or of the rotation periods of either component of the $\alpha$\,Cen system. 
A higher level magnetic activity is thus expected, 
possibly supporting detectable non-thermal coronal radio emission.

To check if the measured radio spectrum of KQ\,Vel is compatible
with non-thermal radio emission from a stellar corona,
we calculated the gyro-synchrotron radio spectrum using a simplified
model for the radio emission from the typical corona surrounding late-type stars.
The model was developed to reproduce the radio spectra of the
active magnetic star members of Algol-type binaries
\citep{umana_etal93,umana_etal99}
and successfully used for the detailed analysis of the radio emission 
of an individual binary of RS\,CVn type
\citep{trigilio_etal01}.
The model assumes the existence of a shell-like structure, the ``halo'',
which is a rough schematization of the extended stellar corona surrounding the late-type star, 
and of a spherical structure, the ``core'',
that represents a more compact region characterized by higher 
magnetic field strength with respect to the average value typical of the halo. 
This could correspond to an active region with an intense 
magnetic field and perhaps the base of a coronal loop filled by mildly relativistic 
non-thermal electrons.
Following \citet{umana_etal93}, the non-thermal electrons are assumed to have
a power-law energy distribution (adopted spectral index $\delta=2$), 
with energies in the range from 10 keV to 5 MeV.
The ``core'' has been assumed to be homogeneously filled by non-thermal electrons;
their density ($n_{\mathrm{rel}}$) is a free-parameter.
In the case of the ``halo", the non thermal electron density was assumed 
to decrease with radial distance, following
$n_{\mathrm{rel}}=n_{\mathrm{rel}}^0 (R_{\ast/}r)^{2}$ \citep{umana_etal99}.
The corresponding electron density at the stellar surface ($n_{\mathrm{rel}}^0$)
is a free-parameter of the model.
The sizes of the two radio emitting regions are defined by
the thickness of the ``halo'' and by the radius of the  ``core''.
Both linear dimensions are also free-parameters.
The magnetic field strengths of these regions
have been assumed to have constant values and are free-parameters.
Those are representative of the average field strengths within these 
coronal regions radiating at the radio regime for the gyro-synchrotron
emission mechanism.
The real magnetic field topology of the coronal regions is expected 
to be highly inhomogeneous, with the magnetic field vectors randomly oriented.
It is then reasonable to assume an average magnetic field vector orientation of 45 degrees
for both core and halo components \citep{umana_etal93}.

The core-halo model requires the definition of six free-parameters for the radio spectrum calculation.
The available multi-frequency radio measurements of KQ\,Vel are only four.
This prevents applying any type of goodness of fit procedure to search for the optimal choice of the model parameters.
We used the core-halo model only to test if plausible parameters exist,
capable of providing a calculated spectrum that fairly matches the observed one.
Further, we have no constraints regarding the real nature
of the hypothesized late-type star.
Consequently, we explored three possible stellar radii for this late-type star.

For the optimal choice of the model parameters,
first we chose the radius of the late-type active stars, then arbitrarily fixed the linear sizes of the core and halo regions.
Finally, we opportunely varied the other two parameters (magnetic field strength and non-thermal electron density) of each 
component to achieve a reasonable visual match between calculated and observed spectra.
The adopted model solution search method is quite easy to perform; in fact each model component 
mainly dominates at a specific spectral range.
In particular, the radio spectrum from the ``core'' dominates at the higher frequencies,
whereas the ``halo'' spectrum dominates the low frequency range.
Hence, the two model parameters of each specific component were varied independently.
Once the visual match in the specific spectral range analyzed was considered satisfactory,  
the corresponding parameters were left fixed, to search for the better parameters of the other model component.
Table~\ref{coha_par} summarizes the model parameters 
that are able to reproduce both the shape and level of the observed radio emission from KQ\,Vel.

The radio spectra corresponding to each component of the core-halo model, for the assumed stellar radius,
have been pictured in three panels of Fig.~\ref{spe_late_star_nonterm_gyro_syn}.
The three different explored sizes of the ``core'' and ``halo'' are recognized by the adopted color code for the calculated 
radio spectrum (small sizes in blue; middle sizes in green; large sizes in red).
The total spectrum has been obtained by combining the calculated spectrum corresponding to
a combination of the free-parameters for a specific model component (core or halo)
with all other spectra calculated using the parameters combinations of the other component.
The total spectra obtained by combining all the core-halo model parameter are pictured in Fig.~\ref{spe_late_star_nonterm_gyro_syn} 
superimposed with the multifrequency radio measurements.
Inspection of the figure indicates that the present analysis cannot significantly constrain the free-parameters.
The only worthwhile result is that the physical parameters used to calculate the 
gyro-synchrotron radio emission from a hypothetical late-type star
are similar to the parameters used to reproduce the radio spectra of other well known
late-type active star members of close binary systems \citep{umana_etal93,umana_etal99,trigilio_etal01}.
This supports the hypothesis that the observed radio emission from KQ\,Vel 
could originate from a late-type companion of the hot, bright, and magnetic Ap star.

\section{Discussion}
\label{sec_discussion}

The analysis of the multi-frequency radio emission from KQ\,Vel showed that 
the measured level and spectral shape  
is compatible with a non-thermal radio spectrum from the corona of
a magnetically active late type star. 
This is a further clue to explain the nature of the KQ\,Vel system.
In fact, the study of KQ\,Vel performed in other spectral bands
suggests the possible existence of a late type companion of the visible early type magnetic Ap star.

High resolution imaging in the near infrared H band, performed by the
PIONIER (Precision Integrated-Optics Near-infrared Imaging ExpeRiment) instrument
at the Very Large Telescope Interferometer (VLTI),
detected an object fainter than the Ap star (difference of magnitudes $\approx 1.8$) and
having an angular separation (18.72 mas) 
compatible with the orbital parameters of the system \citep{scholler_etal20}. 
Further, the TESS 
(Transiting Exoplanet Survey Satellite)\footnote{TESS is a space telescope for NASA's Explorer program, designed to search for extrasolar planets using the transit method.} 
light curve showed clear photometric variability, compatible with a period of $\approx 2.1$ days,
and two flare-like features \citep{scholler_etal20}.
On the other hand, the highly sensitive visible spectra of KQ\,Vel 
show spectral features related to the Ap star only \citep*{bailey_etal15,scholler_etal20}.
On the basis of the expected spectral line detection level,
\citet{scholler_etal20} concluded that the spectral signature of a couple of main sequence F8 type stars (or later) might be unseen.
The fast stellar rotation typical of stars in close binary systems 
(with rotation periods of the order of a day)
could explain the observed photometric variability. 

\begin{figure}
\resizebox{\hsize}{!}{\includegraphics{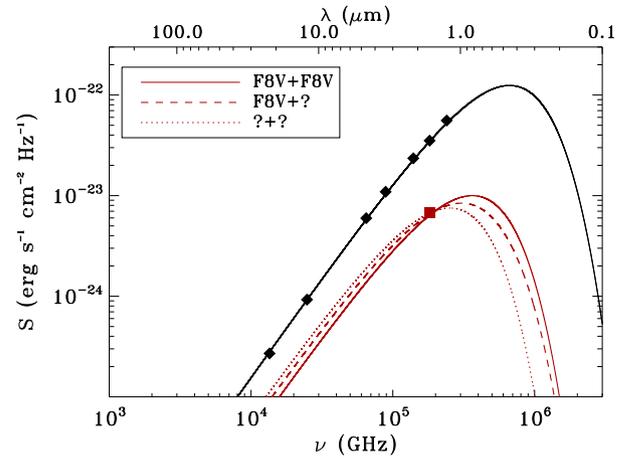}}
\caption{SED of KQ\,Vel. 
The black diamonds are the fluxes of the magnetic Ap star within the 2MASS and the WISE bands.
The red square symbol refers to the flux in the H-band of the close faint object spatially resolved by the
interferometric observations \citep{scholler_etal20}. {Refer to the text for explanation of the line types.}}
\label{fig_sed}
\end{figure}

The magnitudes of KQ\,Vel in the near- and mid- infrared bands are available
in the Two Micron All Sky Survey (2MASS) \citep{2mass}
and the Wide-field Infrared Survey Explorer (WISE) \citep{allwise} catalogues.
The corresponding fluxes in these bands are shown in Fig.~\ref{fig_sed}.
For the magnitude to flux conversion,
the zero-magnitude flux densities of the WISE bands listed by \citet{wright_etal10} have been used;
for the 2MASS bands the fluxes of the zero-magnitudes are instead listed by \citet{cohen_etal03}.

The spectral energy distribution of a black body at the temperature of KQ\,Vel 
has been superimposed on the infrared measurements (black solid line pictured in Fig.~\ref{fig_sed}).
The Stefan-Boltzman relation for the black-body radiation ($L= 4 \pi R^2 \sigma T^4$) allows us to derive the stellar luminosity
following the simple relation:
\begin{displaymath}
\frac {L_{\ast}}{{\mathrm L_{\odot}}}= \left ( \frac{R_{\ast}}{{\mathrm R_{\odot}}}  \right ) ^2 \left ( \frac{T_{\ast}}{{\mathrm T_{\odot}}}  \right ) ^4
\end{displaymath}
Using the stellar parameters listed in Table~\ref{star_par},
the luminosity of the early-type magnetic star of the KQ\,Vel system is about 105 L$_{\odot}$.
That is in good agreement with the value estimated by \citet*{bailey_etal15}.
{The photospheric SEDs of typical Ap stars covering a wide spectral range, typically from the UV up to the NIR domain, 
have been reproduced using models of their stellar atmospheres \citep{shulyak_etal13}.
In this paper, we 
adopted a simple black-body emission, which is only a rough approximation of the real SED
which cannot be assumed to be valid at $\lambda < 1$ $\mu$m.
In any case, the analyzed NIR spectral domain and the high temperature of the Ap star analyzed here
make the differences between the real SED and the black-body emission less dramatic.}

Following the black-body emission framework,
we also analyzed the faint IR companion of the bright Ap star.
Using the typical temperatures and radii ($R_{\ast} \approx 1.3$ R$_{\odot}$;  
$T_{\mathrm{eff}}\approx 6200$ K)
of a main sequence F8 star \citep{eker_etal18},
the combined black body radiation of two F8 stars (as proposed by \citealp{scholler_etal20})
are compatible with the H-band flux of the faint object seen in near infrared interferometry
(red solid line pictured in Fig.~\ref{fig_sed}).
On the other hand, the nature of these stars cannot be unequivocally constrained.
{In fact, stellar parameters compatible with a sub-giant late-type star
also provide black body spectra well in accordance with the H-band data (red dashed line of Fig.~\ref{fig_sed}).
The adopted stellar radius is $R_{\ast}=2$ R$_{\odot}$,
equal to the value used in Sec.~\ref{sec_binary_star}, whereas
the temperature has been adapted to fit the H-band flux   
($T_{\mathrm{eff}}=4300$ K).
Also the black-body emission of two
sub-giant late-type stars alone is enough to reach the measured H-band emission
of the faint IR object (Fig.~\ref{fig_sed}, red dotted line). 
The nature of this possible late-type companion is entirely uncertain,
as the radius and the temperature are degenerate parameters with so few observational constraints.}
Hence, the stellar parameters used to calculate the black body spectrum have to be considered only a plausible combination
able to fit the data, not the true parameters of a possible late-type companion of the Ap star.
Further, the black body emission is only a rough approximation of the real stellar emission,
which dramatically fails to reproduce the true atmospheric emission of the colder stars
dominated by large absorption/emission bands produced by molecular complexes that survive within their cold stellar atmospheres.
The above discussion can be considered only a rough semi-qualitative analysis.
In Appendix~\ref{sed_close_bynaries},
the reliability of the above approach was tested in the case of 
two well studied close binaries composed of late type stars.

Following the clues that suggest the possible existence of a late-type star in the KQ\,Vel system,
we also compared the X-ray and radio emission levels of KQ\,Vel with those typical of late-type stars.
In fact, the radio and X-ray emission of late-type stars characterized by Solar-like magnetic activity
have a well known and fairly stable behavior.
The measured radio luminosity of KQ\,Vel is $L_{\nu, {\mathrm {rad}}}\approx10^{15.6}$ erg s$^{-1}$ Hz$^{-1}$ and
this is compatible with 
the typical radio luminosity (measured at 5 GHz) of the RS\,CVn active binaries, that is
$L_{\nu, {\mathrm {rad}}}=10^{16 \pm 1.5}$ erg s$^{-1}$ Hz$^{-1}$ \citep*{drake_etal89}.
KQ\,Vel is also an X-ray source with an X-ray luminosity $L_{\mathrm X} \approx 10^{30.3}$  erg s$^{-1}$,
which is in good accord with
typical X-ray luminosities of magnetically active close binary systems: $L_{\mathrm X} = 10^{30.5 \pm 1.7}$ erg s$^{-1}$
\citep{drake_etal89,drake_etal92}.
Further, the radio and the X-ray luminosities of stars surrounded by active coronae 
are also correlated, as expressed with the empirical G\"{u}del-Benz relationship \citep{guedel_benz93,benz_guedel94}, which for the 
chromospherically active stars also holds for coherent radio emission \citep{vedentham_etal22}.
In particular, in the case of active binaries (i.e. RS\,CVn type) their radio and X-ray properties
follow the relation $L_{\mathrm X} / L_{\nu, {\mathrm {rad}}} = 10^{14.7 \pm 0.5}$ Hz \citep{benz_guedel94}.
Interestingly, in the case of KQ\,Vel the ratio of its radio and X-ray luminosities is exactly
$L_{\mathrm X} / L_{\nu, {\mathrm {rad}}} \approx 10^{14.7}$ Hz.

{The comparison between the X-ray and radio luminosities of the early-type magnetic stars is largely unexplored.
The ratio $L_{\mathrm X} / L_{\mathrm {\nu, rad}}$ is available in the literature  only for a few  hot magnetic stars.
Some B/A-type magnetic stars have 
$L_{\mathrm X} / L_{\mathrm {\nu, rad}} \approx 10^{12-14}$ Hz \citep{leto_etal17,leto_etal18,leto_etal20,robrade_etal18},
which suggests that the X-ray/radio luminosity ratio of the hot magnetic stars deviates from the GB relation.
This is likely due to the plasma processes occurring within the magnetospheres of the early type magnetic stars,
which differ from those supporting the radio and X-ray emission from the coronae of the active stars.
Extending to the case of KQ\,Vel the X-ray/radio behavior
observed in the few early-type magnetic stars having measured X-ray/radio luminosities ratios, and
using the spectral radio luminosity  theoretically expected of the magnetic Ap star 
($\approx 1.6\times 10^{10}$ erg s$^{-1}$ Hz$^{-1}$, see Sec.~\ref{sec_radio_magnetoshere}), 
we estimate an X-ray luminosity in the range $\approx 10^{22.2}$--$10^{24.2}$ erg s$^{-1}$,
which is many order of magnitudes lower that the measured value ($L_{\mathrm X} \approx 10^{30.3}$  erg s$^{-1}$). 
Therefore, the possible contribution from the magnetic early-type star 
to the measured luminosity ratio of KQ\,Vel is likely negligible.
}

The radio and X-ray properties of KQ\,Vel seem to be well in agreement with
typical behavior of close binary systems characterized by Solar-like magnetic activity. 
On the other hand,
the typical spectral signatures of the chromospheric activity of late-type stars are the CaII  H and K lines observed in emission.
The absence of this signature within the visual spectra of the classical magnetic activity indicators \citep*{bailey_etal15},
the core of the CaII H ($\lambda=3968.469$\,\AA) and K ($\lambda=3933.663$\,\AA) lines observed in emission,
suggests that this hypothetical late-type star could have a ``weak" or ``moderate" \citep{strassmeier_etal88} 
chromospherical magnetic activity.

\begin{figure}
\resizebox{\hsize}{!}{\includegraphics{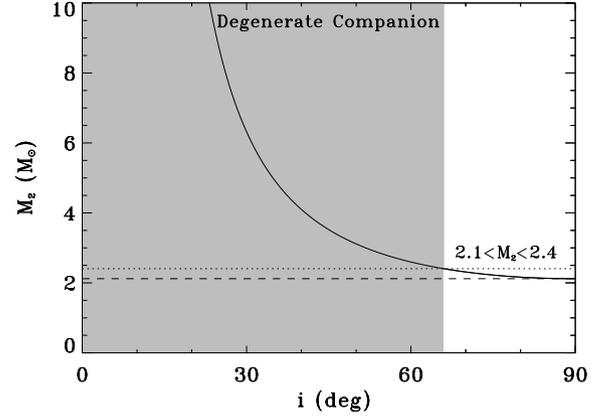}}
\caption{Mass of the faint companion of the Ap star
as a function of the orbital plane inclination.
The solid line {shows the real solutions obtained by numerically solving the Eq.~\ref{eq_mass_fun}.} 
The dashed line refers to the minimum mass compatible 
with the boundary geometric condition corresponding to the orbital plane aligned with the line of sight.
The dotted line refers to the mass value compatible 
with the total mass of a close binary composed of two main sequence F8 type stars (see text).
These masses define the range of allowed total mass of a binary system composed of non-evolved stars.
The dashed area is pictured only to highlight those geometrical conditions of the orbital plane that are not compatible
with the masses of normal stars. More massive stars provide detectable signatures in the visible spectra.
As discussed in the text, a more massive companion should be a degenerate objet.
}
\label{fig_mass_comp}
\end{figure}

\subsection{Considerations on the masses and the inclinations}
\label{sec_orbit_par}

KQ\,Vel is likely a hierarchical triple system, where the faint companion of the early-type magnetic star
could be a close active binary system.
But, the nature of this close binary is basically unknown.
The orbital parameters obtained by fitting the measured radial velocity curve of the Ap star \citep{mathys17}, 
the visible member of the KQ\,Vel system,
places strong constraints on the total mass of the system. 
In particular, the orbital inclination {($i$)} plays a key role 
to estimate indirectly the mass of the companion
{($i$ is the angle between the orbital plane and the plane tangent to the celestial sphere).}

{In the case of a single line spectroscopic binary system, the mass of the unseen component can be constrained
by the binary mass function.
The mass function is derived from Kepler's third law and relates the mass of the star
to orbital information, the mass of the unseen star, and the inclination $i$ of the orbital plane.
Therefore,}
{the total mass of the unseen companion of the KQ\,Vel system can be calculated 
as a function of the orbital plane inclination
using the explicit relation of the mass function:
 \begin{equation}
 \label{eq_mass_fun}
f(M)= \frac {M_2^3 \sin^3i  }{(M_1+M_2)^2  },
\end{equation}
where the adopted value of the mass function \citep{mathys17} is reported in Table~\ref{star_par}.}
Assuming $M_1=3$ M$_{\odot}$ (the mass of the Ap star given by \citealp*{bailey_etal15}),
the corresponding values of $M_2$ are shown in Fig.~\ref{fig_mass_comp} as a function of the inclination $i$.

The lower limit mass ($M_2 = 2.1$ M$_{\odot}$) 
is defined by the geometrical condition of the line of sight lying in the orbital plane ($i=90^{\circ}$).
As previously discussed, 
to satisfy all the observational constraints, the spectral types of the close binary members cannot be earlier than F8V \citep{scholler_etal20}.
The typical mass of an F8V type star is $M_{\ast} \approx 1.2$ M$_{\odot}$ \citep{eker_etal18}.
As a consequence, the upper limit of the total mass of the close binary is $\approx 2.4$ M$_{\odot}$. 
Hence the total mass of the secondary lies in the narrow range $ 2.1 \lessapprox M_2 \lessapprox 2.4$.
On the other hand,
the $M_2$ upper limit indirectly constrains the lower limit of the orbital inclination to $i \approx 65^{\circ}$.
Therefore, if the inclination $i$ of the KQ\,Vel orbital plane were less than $\approx 65^{\circ}$, 
the corresponding total mass of the secondary component 
must be greater than 2.4 M$_{\odot}$ and consequently the non-degenerate
members of the hypothesized close binary are expected to be too bright for their emission level to be
compatible with the measured H band magnitude \citep{scholler_etal20}.
Therefore, the geometrical condition $i \lessapprox 65^{\circ}$ would require the existence of 
a degenerate object possessing a large fraction of the mass of the system to satisfy the orbital parameters of the visible Ap star.

The eccentric orbit of KQ\,Vel suggests that the distance between 
the members of the system is expected {to be} widely variable 
as a function of their orbital motion around the common center of mass.
Note that, only the orbit of the visible star is known.
The projected semi-major axis of the orbit of the visible star is 1.25 AU, listed in Table~\ref{star_par},
corresponding to $\approx 100$ R$_{\ast}$.
Note that this is comparable to the Alfv\'en radius of the magnetic star (see Sec.~\ref{sec_radio_magnetoshere}).
This can be also used to provide a rough estimation of the maximum orbital separation 
between the stellar components, 
about the major axis length ($\approx 2.5$ AU).
At the distance of KQ\,Vel this is about 15.65 mas.
The above rough estimation is close to the separation of 18.72 mas
of the faint IR object measured by \citet{scholler_etal20} using the Very Large Telescope Interferometer.
When the members of the system are closer, i.e. at the distance $\approx 100$ R$_{\ast}$,
the magnetic field strength of the bright magnetic star is $\approx 7.5$ mG, 
obtained using the inverse cube law that describes a simple dipole with polar strength 
$B_{\mathrm p}\approx 7.5$ kG \citep*{bailey_etal15}.
Hence the magnetic field strength in the spatial region where the unknown companion orbits
is not negligible, furthermore, this is also expected to be strongly variable as the stars revolve around the center of mass
along their eccentric orbits.

\begin{figure}
\resizebox{\hsize}{!}{\includegraphics{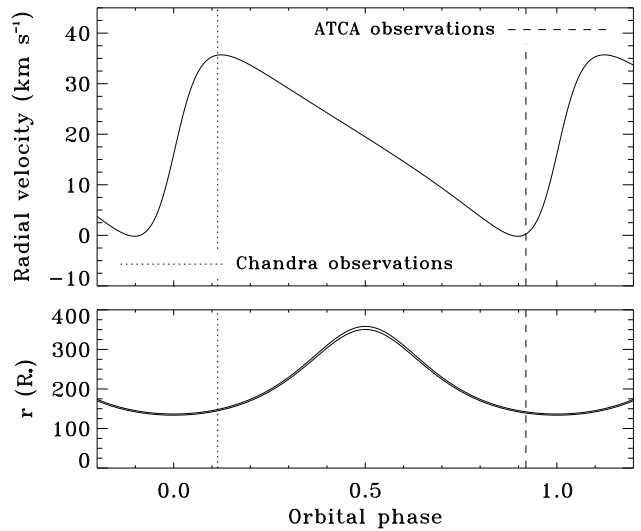}}
\caption{{Top panel: radial} velocity curve of the bright Ap star member of the KQ\,Vel system.
The adopted orbital parameters are listed in Table~\ref{star_par}.
The vertical dashed line located the orbital phase when the ATCA observations were performed.
{Bottom panel: distance between the components as a function of the orbital phase.}}
\label{fig_rv_kqvel}
\end{figure}

The orbital position of the magnetic early type star during the ATCA radio observations
can be calculated using the ephemeris  \citep{mathys17}  listed in Table~\ref{star_par}. 
The radial velocity curve of KQ\,Vel, pictured in {the top panel of} Fig.~\ref{fig_rv_kqvel}, 
has been calculated using the orbital parameters provided by \citet{mathys17}.
The calculation has been performed using the procedure {\sc helio\_rv}  enabled within the {\sc idl} data language,
which returns the heliocentric radial velocity of a spectroscopic binary with known orbital parameters. 
The orbital phase of the magnetic star at the time the ATCA observations were performed is indicated 
by the vertical dashed line in Fig.~\ref{fig_rv_kqvel}.

The distance between the components of the KQ\,Vel stellar system
is expected to be change substantially as they orbit around the common center of mass.
{
The distance can be estimated by adding the radial distances ($r$) of the components 
from the focus of the ellipse where the common center of mass is located.
The polar equation of the orbit of each component is:
\begin{displaymath}
r_{\mathrm i}=\frac{a_{\mathrm i} (1-e^2)}{1+e \cos \theta},
\end{displaymath}
where  $\theta$ is the polar angle with the origin coinciding with the periastron. 
Both stars orbit around the common center of mass with the same eccentricity $e$, listed in Table~\ref{star_par}.
The semi-major axis of the unseen component can be derived by using the definition of center of mass:
$a_2=a_1 \times M_1/M_{2}$,
where $M_2$ is constrained in the range 2.1--2.4 M$_{\odot}$, as discussed above.
The distance $r=r_1+r_2$ 
between the components of the KQ\,Vel system as a function of the orbital phase is reported in the bottom panel
of Fig.~\ref{fig_rv_kqvel}.
The two almost indistinguishable lines refer to the two orbital inclination angles taken into account ($i=90^{\circ}$ and $i=65^{\circ}$).

The widely variable distance between the components of the KQ\,Vel system 
(range $\approx130$--360 R$_{\ast}$, lower limit close to the minimum component separation roughly estimated above)
implies that it
}
is likely to expect strongly variable magnetic interaction as a function of orbital position.
{Further, as suggested by the behavior at the low frequency
side of the simulated spectra, pictured in Fig.~\ref{spe_late_star_nonterm_gyro_syn},
the size of the largest coronal component contributing to 
radio emission cannot be constrained by the radio measurements reported here
(i.e. high sensitivity measurements at frequencies close to or lower than $\approx 1$ GHz might be helpful).} 
If the observed radio emission originates from the 
magnetic corona of a late-type star,
magnetospheric interaction may also affect the physical mechanisms able to produce non-thermal electron acceleration.
This is only a speculative hypothesis;
long-term monitoring of KQ\,Vel at the radio regime could be useful to confirm or refute the idea.

Finally, the age of the KQ\,Vel system (260 Myr)
puts strong constraints on the nature of the late-type stars of the active close binary, 
hypothesized to be unknown companion to the bright Ap star.
These cannot be evolved stars. The close binary is likely composed of two solar-like young stars in close orbit,
like the case of  $\sigma^2$\,CrB. 
This is a young close binary, about 10 Myr old \citep{strassmeier_rice03}, of RS\,CVn type.
This is a well studied active binary
composed of two main sequence F9V and G0V dwarf stars (see Appendix~\ref{sed_sigma2crb} for additional details regarding $\sigma^2$\,CrB).

It is worth noting that the rotation axis of the magnetic Ap star, i.e. the dominant member in the visible domain of the KQ\,Vel system, 
is tilted by only $\approx 16^{\circ}$ with respect to the line of sight. 
It follows, that in the case of large inclination of the orbital plane, 
the axial tilt between the rotation and the orbital axes (obliquity) is consequently expected to be large. 
The obliquity upper limit is $\approx 74^{\circ}$, in the case of the orbital plane being aligned with the line of sight ($i=90^{\circ}$). 
It follows, that in this case limit the Ap star revolves around the center of mass with its rotation axis almost contained within the orbital plane,
this would be a really strange configuration.

\subsection{Revisiting the neutron star companion scenario}
\label{sec_accret_ns}

{
The lower limit mass (2.1 M$_{\odot}$, see previous section) of the companion
to the Ap star in the KQ\,Vel system is compatible with the possible presence of a compact object,
such as e.g. a NS \citep*{bailey_etal15}. 
A NS companion can accrete matter from the wind sprayed out by the magnetic Ap star.
In the case of a magnetized NS, the accretion of matter could also be a possible source of synchrotron radio emission.
The accreting NS scenario has been considered by \citet{oskinova_etal20} to explain the X-ray emission of KQ\,Vel,
but ruled out above because the very low ambient plasma density 
predicts an X-ray emission level
three orders of magnitudes lower than the measured X-ray luminosity.

The possible effects induced by the magnetosphere of the Ap star depend on
the component separation, which is a function of the orbital phase. 
The measurements of the X-ray emission from KQ\,Vel have been performed by the {\it Chandra} X-ray telescope
on 2016-Aug-20 (mean UT time about 11:00).
The orbital phase when the X-ray measurements was performed is marked in Fig.~\ref{fig_rv_kqvel}
by the vertical dotted line, which corresponds 
to a distance between the components of
$\approx 148$ R$_{\ast}$. 
As discussed in Sec.~\ref{sec_radio_magnetoshere},
the magnetic Ap star is the source of a weak wind;
two recipes have been taken into account to estimate its properties.
Using the two different values of the wind mass loss rates adopted in this paper, 
we estimated the ambient density where the hypothesized NS is located
at the epoch of the X-ray measurement.
In stars with centrifugal magnetospheres accumulation of matter is expected within their magnetospheres.
But in cases of dynamical magnetospheres, which is the case we're dealing with (see Sec.~\ref{sec_radio_magnetoshere}),
the accumulation of plasma is balanced by the gravitational infall, making reasonable to roughly assume 
the mass flux inside the magnetosphere coinciding with the flux of mass lost from the whole stellar surface due to the wind.
In the case of a lower wind mass loss rate ($\dot{M} \approx 10^{-14.3}$ M$_{\odot}$ yr$^{-1}$),
the corresponding Alfv\'en radius ($R_{\mathrm A} \approx 415$ R$_{\ast}$) is larger that
the component separation ($r\approx 148$ R$_{\ast}$).
Making the simplest assumption of a density profile $n \propto \dot{M}/(4\pi v_{\infty} r^2)$, 
which is justified by the condition of dynamical magnetosphere,
the ambient electron density $n_{\mathrm e}\approx 0.39$ cm$^{-3}$ was derived.
The other wind regime taken into account in this paper 
is $\dot{M} \approx 10^{-11.9}$ M$_{\odot}$ yr$^{-1}$,
which is able to open the magnetosphere at $R_{\mathrm A} \approx 106$ R$_{\ast}$.
The above value of the Alfv\'en radius is lower than the component separation. 
For the local electron density calculation, we crudely estimate the 
mass lost from the two polar caps where the magnetic field lines are assumed to be open 
\citep*{ud-doula_etal02,ud-doula_etal08,petit_etal17}.
To do this, we calculate the  fraction of the stellar surface where the wind can freely propagate.
This is the ratio between the polar caps area calculated using the latitude where the last closed magnetic field line, 
assumed to be a simple dipole,
crosses the stellar surface, and the total stellar surface.
In the case of $R_{\mathrm A} = 106$ R$_{\ast}$ and $R_{\ast}=2.7$ R$_{\odot}$,
the fraction is $\approx 0.5$\%, the corresponding actual wind mass loss rate is 
$\dot{M}_{\mathrm {act}} \approx 10^{-14.2}$ M$_{\odot}$ yr$^{-1}$,
which corresponds to a local ambient electron density of about  0.46 cm$^{-3}$
at the the distance of 148 R$_{\ast}$.
We emphasize that the two different mass loss
recipes, when magnetic confinement is accounted for, result in similar
densities at the companion's position.

Adopting the same parameters as in \citet{oskinova_etal20}, and assuming a spherical wind density radial profile,
the corresponding ambient electron density is $n_{\mathrm e}\approx 1.6$ cm$^{-3}$,
which is higher than both values estimated in this paper.
Therefore, the wind parameters given in this paper provide a more stringent condition regarding 
the capability of an hypothetical accreting NS to support the observed X-ray luminosity.
The inability to reproduce one (the X-rays) of the two observables (X-rays and radio) analyzed in this paper
convinced us to not investigate further the accreting NS scenario.
}

\section{Summary and conclusions}
\label{sec_conclusion}

In this paper the multi-frequency radio detection, performed by the ATCA interferometer, 
of the KQ\,Vel multiple stellar system is reported.
These new radio measurements have been used to put constraints on the nature of this enigmatic stellar system.
When observed at visible wavelengths high-resolution spectra of KQ\,Vel 
show only evidence of an early-type magnetic star \citep*{bailey_etal15}.
This long-period magnetic star ($P_{\mathrm {rot}} = 2830$ d; \citealp{giarrusso_etal22})
is a member of a high eccentricity multiple system as directly evidenced by the measured radial velocities \citep{mathys17}, 
but no evidence of the companion is seen at visible wavelengths, hence its nature remains unclear.
The only direct measurement of the companion's brightness was performed at the infrared domain 
with interferometric observations reported by \citet{scholler_etal20}.

The stellar parameters of the visible magnetic star
belonging to the KQ\,Vel system are well known, 
which enables us to check the possible magnetospheric origin of the measured radio emission.
Comparing the radio luminosity of KQ\,Vel with the luminosity level predicted
by the universal law of radio emission from well ordered and stable co-rotating magnetospheres,
that was empirically discovered by \citet{leto_etal21}, confirmed by \citet{shultz_etal22}, 
and then theoretically supported by \citet{owocki_etal22},
we found that the observations are dramatically above ($\approx 5$ dex) the theoretical prediction,
because the star is an extremely slow rotator.
We therefore reasonably ruled out non-thermal radio emission from the magnetosphere of the 
visible early-type magnetic star as a possible origin of the measured radio emission.

KQ\,Vel is also a bright X-ray source.
To explain the X-ray emission level and spectrum, 
\citet{oskinova_etal20} proposed a scenario where the X-ray emission originates from a hot plasma
shell surrounding an unseen degenerate magnetic companion, probably a neutron star.
We analyzed if the above scenario, able to explain the X-ray emission,
is also able to explain the origin of the radio emission.
The radio emission level of the X-ray emitting thermal plasma (given by the thermal bremsstrahlung emission mechanism) 
is too low to explain the measured radio emission.
On the other hand, a significant fraction of the hot plasma gravitationally captured by the neutron star
might be accelerated to relativistic energies. 
We also checked if the non-thermal electrons falling inside the deep magnetospheric
regions close to a magnetic degenerate object 
are able to produce detectable radio emission by the gyro-synchrotron emission mechanism.
The calculated theoretical emission level is close to the observed one at the higher frequencies.
However,
we found that the low frequency side of the calculated radio spectrum
suffers significantly from frequency-dependent absorption effects produced by the thermal plasma
envelope (responsible for the X-rays) surrounding the magnetospheric regions responsible for the non-thermal radio emission.
Therefore neither thermal nor non-thermal emission from a propelling neutron star are consistent with the properties of KQ\,Vel's radio spectrum.

\citet{oskinova_etal20} also considered the possibility of an active binary star companion.
This was corroborated by the discovery of
periodic photometric variability, with a period of about 2.1 days, 
evidence of flares, and the interferometric detection of a companion star in the right magnitude range for a solar-type object.
This suggests the possible existence of a companion with Solar-like magnetic activity \citep{scholler_etal20}, 
that might be responsible for the observed photometric behavior.
Hence, KQ\,Vel could be a hierarchical multiple stellar system 
hosting a late-type star (or stars) with an extended corona sustained by a large-scale magnetic field generated 
by the dynamo mechanism acting within the deep convective layers below the stellar photosphere
(i.e. the active members of close binaries like the RS\,CVn systems).
In principle, the coronal magnetic activity of such a possible late-type companion might explain both the X-ray and radio emission of KQ\,Vel.
Further, the measured radio and X-ray luminosity ratio is well in accordance with the G\"{u}del-Benz law prediction in the case of an RS\,CVn system.
We calculated the non-thermal radio spectrum arising from an extended stellar corona.
For a hypothesized active member of the KQ\,Vel system,
we adopted plausible stellar parameters
that are compatibile with those used to reproduce the observed radio spectra
of known active binary systems well studied at radio wavelengths.
Using model parameters that are canonical for active close binary systems,
the calculated radio spectrum, scaled at the distance of KQ\,Vel, matches the observed one.

The scenario where the KQ\,Vel system hosts a star characterized by Solar-like magnetic activity seems preferred.
On the other hand, the mass of the early-type magnetic star and its orbital parameters 
places strong constraints on the mass of the unknown companion.
In fact, the minimum mass of the companion that is compatible with the mass function of KQ\,Vel
is higher than 2.1 M$_{\odot}$, which is in accordance with the total mass of typical 
RV\,CVn active binary system.
But the inclination of the orbital plane of the KQ\,Vel system is unknown.
The above reported lower limit of the mass of the unseen companion 
was obtained in the case of the orbital plane perfectly aligned with the line of sight.
If the orbital plane has a {smaller inclination (i.e. the orbit view becomes close to the face-on view)}, 
the mass of the companion required to explain the observed radial velocity curve increases.
If the required mass of the unknown companion of the KQ\,Vel system needs to increase,
the star will be consequently brighter.
\citet{scholler_etal20} estimated that the spectral signatures  of a couple of normal F8 type stars
are undetectable in the high-sensitivity visual spectra of KQ\,Vel.
The above condition definitively fixes the upper limit of the total mass of the KQ\,Vel companion in the case of normal 
main sequence stars.
To take into account possible {lower} 
inclination of the orbital plane,
the existence of a degenerate object is mandatory. 
{The eccentric orbit implies that along its orbital motion the unseen companion meets widely variable conditions
(i.e. the magnetic field strength and the plasma density provided by the magnetic Ap star).
Such variable conditions may affect the X-ray emission and likewise the radio emission.}
This motivates interferometric monitoring, 
which together with existing RVs would constrain the inclination of the orbit and therefore provide a definitive test.
The unknown component of the KQ\,Vel system could be
an exotic close binary consisting of a degenerate object and a late-type star \citep{oskinova_etal20},
possibly characterized by enhanced solar-like magnetic activity that explains the radio and X-ray observing features.

The measured radio flux density of KQ\,Vel is quite low ($\approx 100$ $\mu$Jy),
this makes it hard to study the radio source morphology and also to reveal possible radio emission variability related to the orbital position.
Indeed, it is well known that the non-thermal radio emission of active binary systems is strongly variable.
Future high-sensitivity radio measurements with higher spatial resolution, 
performed using the forthcoming SKA or its precursors (MeerKAT, ASKAP, or ngVLA),
will be crucial for definitively revealing the nature of the enigmatic KQ\,Vel stellar system.

\section*{Acknowledgments}

We thank the anonymous referee for his/her very useful comments and criticisms
that allowed us to significantly improve the paper.
This work has extensively used the NASA's Astrophysics Data System, and the 
SIMBAD database, operated at CDS, Strasbourg, France. 
LMO acknowledges support from
the DLR under grant FKZ\,50\,OR\,1809. 
MES acknowledges financial support from the Annie Jump Cannon Fellowship, supported by the University of Delaware and endowed by the Mount Cuba Astronomical Observatory.
RI acknowledges funding support for this research from a grant by the National Science Foundation, AST-2009412.
MG acknowledges financial contribution from the agreement ASI-INAF n.2018-16-HH.0.
FL was supported by ``Programma ricerca di ateneo UNICT 2020-22 linea 2''.
A special thank to the project MOSAICo 
(Metodologie Open Source per l'Automazione Industriale e delle procedure di CalcOlo in astrofisica) funded by Italian MISE (Ministero Sviluppo Economico).
Finally, we thank Tim Bedding who shared with us the TESS light curve.

\section*{Data availability}
The data underlying this article are available within the body of the paper and within its tables.



\appendix

\section{The IR SEDs of two close binaries}
\label{sed_close_bynaries}

The black-body spectrum is only a
rough approximation of the true spectral energy distribution of the photospheric emission of stars,
in particular late type stars.
This approximation has been adopted here only to check 
if the expected emission 
from stars belonging to close binary systems classified as RS\,CVn type, that also have well known
sizes and temperatures,
might be considered fairly in accordance with measurements in the infra-red spectral domain.
In the following, two well studied close binary of RS\,CVn type have been taken into account.
The two selected binaries have almost similar total masses (2.2 and 2.4 M$_{\odot}$,
compatible with the mass constraint of the secondary component of the KQ\,Vel system), 
but with different evolutionary stages.
The young $\sigma^2$\,CrB is composed of a pair of similar dwarf stars, whereas 
TY\,Pyx is composed of two sub-giant stars of the same spectral type.

\subsection{The young active binary $\sigma^2$\,CrB}
\label{sed_sigma2crb}

\begin{figure}
\resizebox{\hsize}{!}{\includegraphics{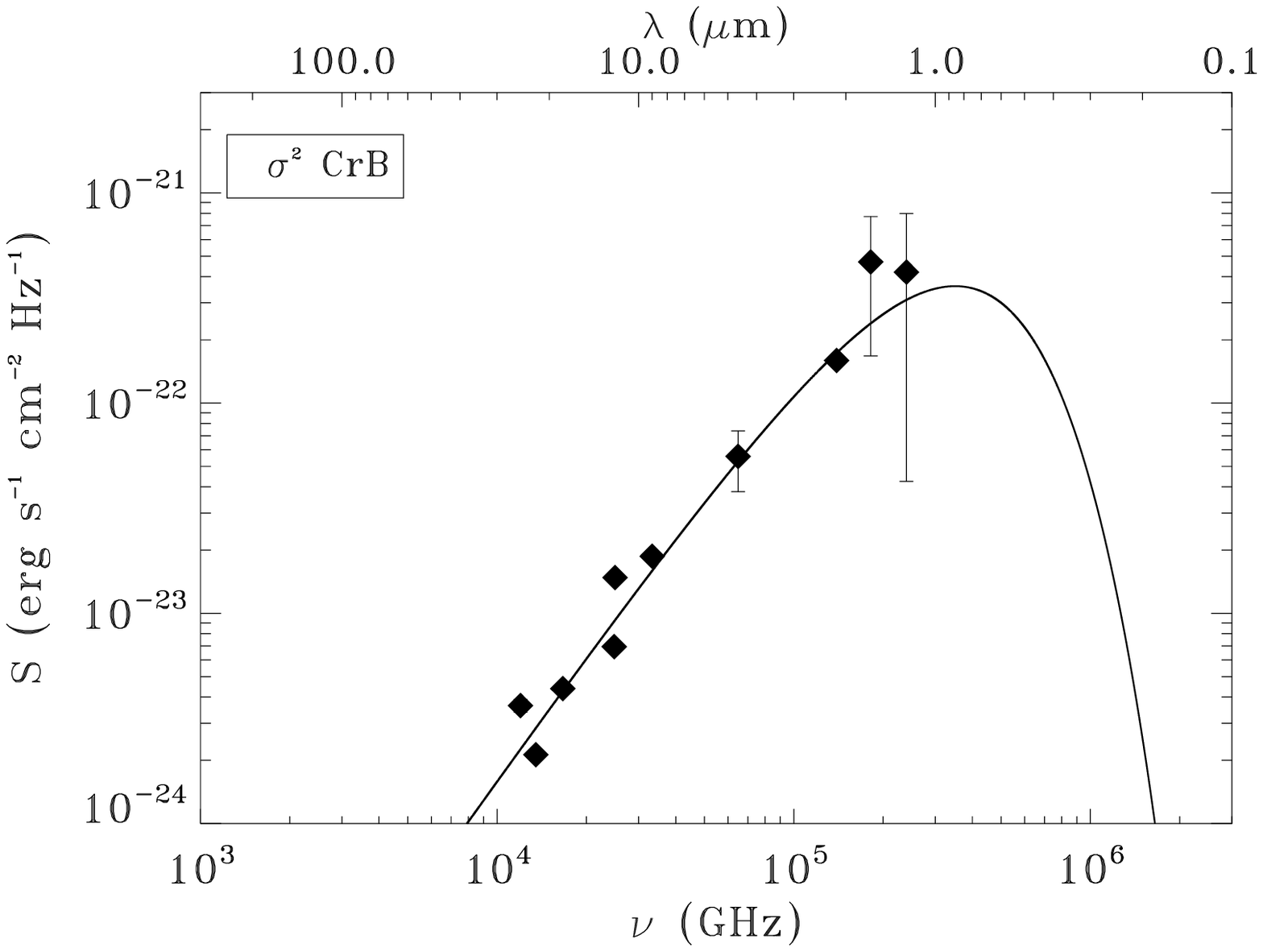}}
\resizebox{\hsize}{!}{\includegraphics{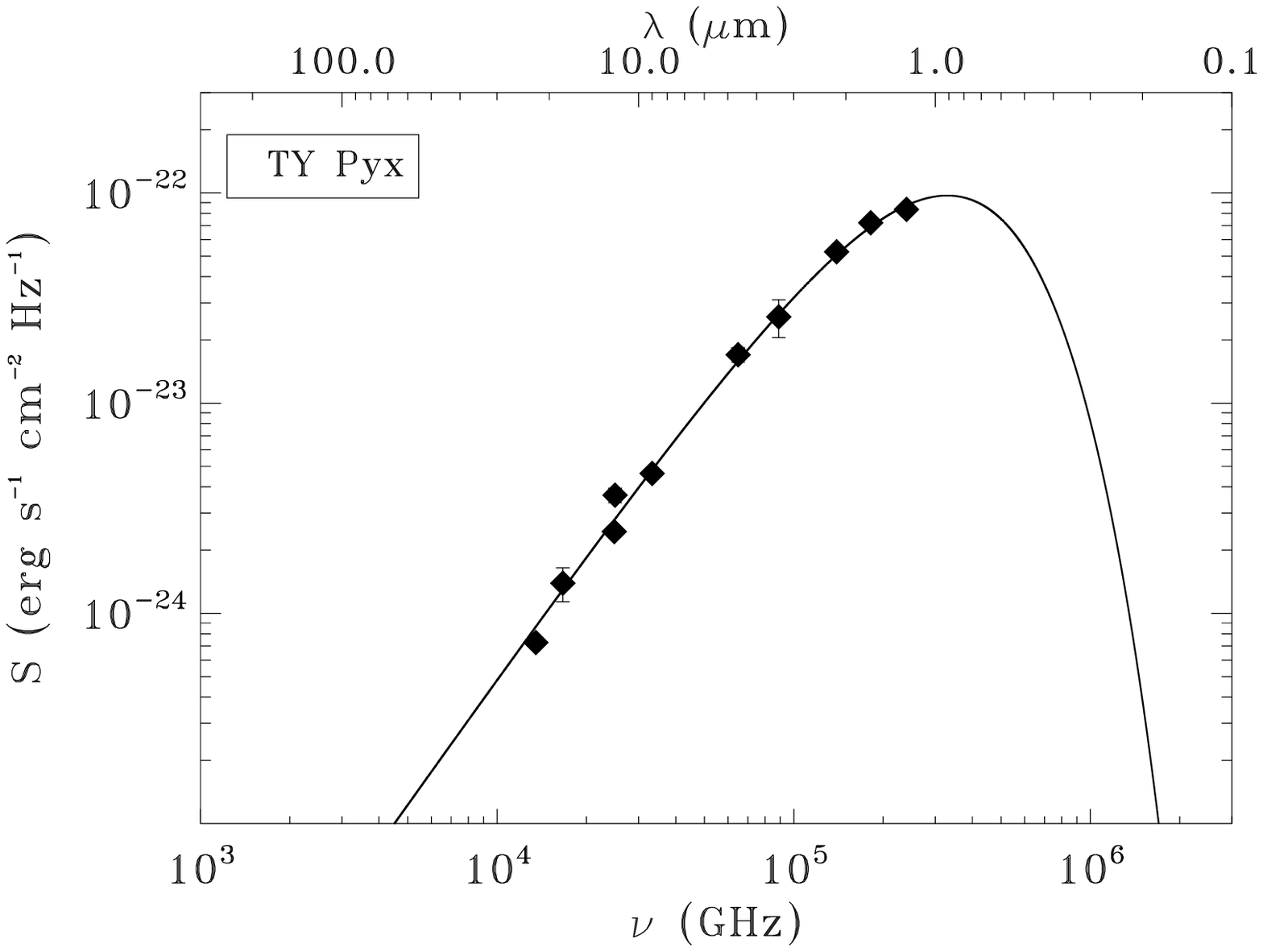}}
\caption{SEDs of two well studied close binaries of RS\,CVn type.
The black diamonds are the fluxes of $\sigma^2$\,CrB  and TY\,Pyx within the 2MASS, WISE, IRAS, and AKARI bands.
Top panel: SED of $\sigma^2$\,CrB,
the solid line is the combined black body emissions of two F9V and G0V main sequence stars at the distance of  $\sigma^2$\,CrB.
Bottom panel: SED of TY\,Pyx,
the solid line is the combined black body emissions of two sub-giant stars of G5IV-type at the distance of  TY\,Pyx.
}
\label{fig_sed_two_rscvns}
\end{figure}

The close binary $\sigma^2$\,CrB is composed by two main sequence solar like stars
(both masses $\approx 1.1$ M$_{\odot}$),
with spectral types respectively F9V and G0V. Their evolutionary tracks  
clearly indicate both are located very close to the zero age main sequence
with ages of $\approx 10$ Myr  \citep{strassmeier_rice03}.  
This is classified as a variable of RS CVn type
and is close to the Earth, at a distance $21.1\pm 0.5$ pc \citep{hipparcos}.
The  detection  of its incoherent non-thermal radio emission at the microwave frequency range  (5 GHz)  
was first reported by \citet{spangler_etal77}.
$\sigma^2$ CrB also shows
evidence of flares, typical of the cromospherically active binaries, 
at the X-ray, UV, and radio bands.
The measured radio emission at $\nu=8.46$ GHz
rises from a quiescent level of $\approx 1$ mJy up to beyond 100 mJy, at the flare maximum \citep{osten_etal00}.
Scaling at the stellar distance, the corresponding quiescent and flaring 
spectral radio luminosities are $L_{\nu,{\mathrm {rad}}}\approx 2.7\times10^{15}$ and 
$L_{\nu,{\mathrm {rad}}}\approx 3\times10^{17}$ erg s$^{-1}$ Hz$^{-1}$.
The X-ray luminosities reported in the ROSAT and XMM catalogues are respectively: $L_{\mathrm X} \approx 4\times10^{30}$ 
and $\approx 5\times10^{30}$ erg s$^{-1}$ \citep{hinkel_etal17}.
Further, this close binary showed clear evidence of coherent emission at low radio frequency (144 MHz)
\citep{toet_etal21}.
Both stars have synchronized rotation periods of 1.157 days \citep{strassmeier_rice03}.
The dwarf stars members 
of this close binary rotate significantly faster than the Sun (period $\approx 27$ days).
This explains their higher activity level, in accordance with the correlation between the period and non-thermal
radio luminosity \citep{mutel_lestrade85}.

In the top panel of Fig.~\ref{fig_sed_two_rscvns} the 2MASS \citep{2mass}, WISE \citep{allwise}, IRAS \citep{iras_faint_v2}, 
and AKARI \citep{akari_irc} reliable
measurements of $\sigma^2$\,CrB, with the corresponding errors, are reported.
The stellar radius and temperature of the primary (the F9V star) are: $R_{\ast}=1.14 \pm0.04$ R$_{\odot}$ and
$T_{\mathrm {eff}}=6000 \pm 50$ K, the corresponding stellar parameters of the secondary (the G0V star)
are: $R_{\ast}=1.10 \pm0.04$ R$_{\odot}$ and
$T_{\mathrm {eff}}=5900 \pm 50$ K \citep{strassmeier_rice03}.
The combined black-body emission, calculated using radii and temperatures reported above, 
is also pictured in top panel of Fig.~\ref{fig_sed_two_rscvns}. 
Looking at the figure, it is evident that 
the measured fluxes of $\sigma^2$ CrB  are fairly in accordance with the
calculated black-body emission of both stellar components.

\subsection{The evolved active binary TY\,Pyx}
\label{sed_typyx}

TY\,Pyx is a well known active binary of RS\,CVn type.
This binary is in the Solar neighborhood,
the system is $54.60 \pm 0.05$ pc away from Earth \citep{gaia_dr3}.
The age of TY\,Pyx is 3.9 Gyr \citep{holmberg_etal09} and it is comparable with the Sun's age.
The stellar components are both sub-giants of G5IV-type \citep{strassmeier_etal88}, 
with a common rotation period of $\approx 3.2$ days \citep{andersen_popper75}.
These stars are slightly more evolved than the Sun probably because they are more massive.
Both components have similar masses, each $\approx 1.2$ M$_{\odot}$ \citep{andersen_popper75}.
The detection of non-thermal radio emission was firstly reported by \citet{owen_gibson78}.
This binary is also a well studied X-ray source \citep{franciosini_etal03}.

In the bottom panel of Fig.~\ref{fig_sed_two_rscvns} the 2MASS \citep{2mass}, WISE \citep{allwise}, IRAS \citep{iras_faint_v2}, 
and AKARI \citep{akari_irc} reliable
measurements of TY Pyx, with the corresponding errors, are reported.
The components of the TY\,Pyx system have approximatively equal radii and temperatures: $R_{\ast}\approx 1.65$ R$_{\odot}$ and
$T_{\mathrm {eff}} \approx 5600$ K \citep{andersen_popper75}.
The combined black-body emission of two equal G5IV stars, calculated using radii and temperatures reported above, 
is also pictured in the bottom panel of Fig.~\ref{fig_sed_two_rscvns}. 
Looking at the figure, it is evident that 
the measured fluxes of TY Pyx  are in accordance with the
combined black-body emission of both stellar components.

\end{document}